\documentclass[useAMS,usenatbib,usedcolumn]{mn2e}

\def\HI{\ifmmode{\rm HI}\else{H\/{\sc i}}\fi}

\def\Ha{\hbox{\rm H$\alpha$}\ }

\def\lsun{\ifmmode{{\mathrm L}_{\odot}}\else{L$_{\odot}$}\fi} 

\def\deg{\hbox{$^\circ$}}
\def\arcmin{\hbox{$^\prime$}}
\def\arcsec{\hbox{$^{\prime\prime}$}}

\def\magasas{\nobreak\mbox{$\;$mag$\,$arcsec$^{-2}$}}
\def\msun{\ifmmode{{\mathrm M}_{\odot}}\else{M$_{\odot}$}\fi} 
\def\msunpc2{\ifmmode{{\mathrm M}_{\odot} \, {\mathrm{pc}}^{-2}}\else{M$_{\odot} \, {\mathrm {pc}}^{-2}$}\fi}
\def\keer{{\small $\times$}}

\newcommand{\NII}{[N\/{\sc{ii}}]}

\usepackage[figuresright]{rotating}
\usepackage{lscape}
\usepackage{graphics}
\usepackage{epsfig}
\usepackage{multirow}
\usepackage{bigdelim}
\usepackage{bigstrut}

\defcitealias{Noordermeer05}{Paper~I}

\title[The stellar mass distribution in early-type disk galaxies]{The stellar
  mass distribution in early-type disk galaxies: surface photometry and
  bulge-disk decompositions}
\author[E.~Noordermeer \& J.~M.~van der
  Hulst]{E.~Noordermeer,$^{1,2}$\thanks{email: edo.noordermeer@nottingham.ac.uk}
  and J.~M.~van der Hulst$^1$ \\ 
  $^1$Kapteyn Astronomical Institute, University of Groningen, PO Box 800,
  9700 AV Groningen, The Netherlands \\
  $^2$University of Nottingham, School of Physics and Astronomy, University
  Park, NG7 2RD Nottingham, UK}
\begin{document}

\date{accepted for publication in MNRAS, 19-12-2006}

\maketitle

\begin{abstract}
We present deep B- and R-band surface photometry for a sample of 21 early-type 
disk galaxies with morphological types between S0 and Sab and absolute B-band
magnitudes between -17 and -22. Six galaxies were also observed in I. We
present radial profiles of surface brightness, colour, ellipticity, position
angle and deviations of axisymmetry for all galaxies, as well as isophotal and
effective radii and total magnitudes. We have decomposed the images into
contributions from a spheroidal bulge with a general S\'ersic profile and a
flat disk with an arbitrary intensity distribution, using an interactive, 2D
decomposition technique. We caution against the use of simple 1D decomposition
methods and show that they can lead to systematic biases, particularly in the
derived bulge parameters.  

We study in detail the relations between various bulge and disk parameters. 
In particular, we find that the bulges of our galaxies have surface brightness
profiles ranging from exponential to De Vaucouleurs, with the average value of
the S\'ersic shape parameter $n$ being 2.5. In agreement with previous
studies, we find that the shape of the bulge intensity distribution depends on
luminosity, with the more luminous bulges having more centrally peaked light
profiles (i.e.\ higher $n$). 
By comparing the ellipticity of the isophotes in the bulges to those in the
outer, disk dominated regions, we are able to derive the intrinsic axis ratio
$q_b$ of the bulges. The average axis ratio is 0.55, with an rms spread of
0.12. None of the bulges in our sample is spherical, whereas in some cases,
the bulges can be as flat as $q_b = 0.3 - 0.4$. The bulge flattening seems to
be weakly coupled to luminosity, more luminous bulges being on average
slightly more flattened than their lower-luminosity counterparts. 
Our finding that most bulges are significantly flattened and have an intensity
profile shallower than $R^{1/4}$ suggests that `pseudobulges', formed from
disk material by secular processes, do not only occur in late-type spiral
galaxies, but are a common feature in early-type disk galaxies as well. 
  
Most galaxies in our sample have radial colour gradients, becoming bluer
towards larger radii. Although this can partly be explained by the radially
declining contribution of the red bulges to the observed light, we show that
disks must also have intrinsic colour gradients. 
\end{abstract}


\section{Introduction}
\label{sec:introduction}
One of the outstanding problems in our understanding of the structure
and evolution of galaxies is the r\^ole played by dark matter. 
Paradoxically, however, one of the main obstacles towards an exact measurement
of the amount and distribution of dark matter is, in many cases, formed by our
limited knowledge of the contribution of the stars to the gravitational
potential.   
The contribution of the stellar component is expected to be
particularly important in early-type disk galaxies. 
Early-type disk galaxies are among the brightest galaxies in the
universe, both in terms of total luminosity and of surface brightness
\citep{Roberts94}.   
It is to be expected that in these galaxies, the stellar mass is
dynamically much more important than in low luminosity and
low surface brightness galaxies, which are generally believed to be
dark matter dominated \citep[e.g.][]{Carignan88, DeBlok97,
Swatersthesis, Cote00}. 

Historically, spiral galaxies were considered as a combination of a
more or less spherical bulge and a flattened disk. 
Bulges were normally modelled as miniature elliptical galaxies, with
surface brightness following an $R^{1/4}$-law
\citep[following][]{DeVaucouleurs48, DeVaucouleurs58}, whereas the
azimuthally averaged disk brightness was usually described with an
exponential profile \citep{DeVaucouleurs59, Freeman70}.   
 
It was readily noted that not all galactic disks are well described by
simple exponential profiles at all radii \citep{Freeman70,
VanderKruit79}, and recent developments have shown that many bulges
are not the simple, structureless spherical bodies as previously
thought either.   
HST observations by \citet{Carollo97, Carollo98} of the cores of 75
spiral galaxies revealed a wealth of nuclear structure, including
nuclear spirals, rings and dust lanes.   
Many of their galaxies showed signs of nuclear star formation.  
\citet{Erwin99,Erwin02} and \citet{Erwin03} showed that many galaxies
have nuclear bars and other disk-like structures in their centres. 
They noted that the central light concentrations in many spiral
galaxies seem to originate from highly flattened structures, rather
than from spherical bodies. 
The photometric profiles of many spiral galaxy bulges are described
better by exponential or $R^{1/2}$-profiles than by the classical
$R^{1/4}$-law \citep{Andredakis94, Andredakis95, DeJong96,
Courteau96, Carollo99}.    
Finally, spectroscopic observations show that many bulges are much
more rotationally supported than elliptical galaxies
(\citealt{Illingworth82, Kormendy82, Kormendy93}; see also the recent 
results from the SAURON project for an overview of the complex
dynamics, including disk-like rotation, in the centres of many spiral
galaxies: \citealt{Emsellem04, Fathi04}). 

All these observations led \citet{Kormendy04} to postulate that many
central light concentrations are not classical, spheroidal bulges with
a similar formation history as ellipticals, but rather disk-like
structures, formed by secular processes; they call these disk-like
bulges `pseudobulges'. 

For a study of the global dynamics of spiral galaxies, a proper
understanding of the structure of bulges and disks is crucial. 
Irrespective of the exact formation mechanism of (pseudo)bulges, they
will have experienced a different evolutionary history than their
surrounding disks. 
As a result, their stellar populations, and corresponding
mass-to-light ratios, are expected to be different. 
Accurate bulge-disk decompositions are therefore needed to determine
the contribution of each component to the gravitational potential. 
Furthermore, the vertical structure of the bulge has a strong
influence on the peak rotation velocity in a galaxy: a flat, disk-like
structure will generally have a higher circular rotation speed in the
plane than a spherical system with the same projected surface density
(see e.g.\ figure~2-12 in \citealt{Binney87}).

The current paper is part of a larger study of the relation between
dark and luminous matter in nearby, early-type disk galaxies.  
In a previous paper \citep[][ hereafter paper~I]{Noordermeer05}, we
presented \HI\ observations of a sample of 68 such galaxies. 
In an accompanying paper to the current one \citep{Noordermeer06b}, we present
rotation curves for a subset of the galaxies from \citetalias{Noordermeer05},
and in a future publication, we will compare them to the rotation velocities
expected from the visible matter in order to infer the dark matter
distribution.   
Here, we present another ingredient for our study, a study of the
dynamical impact of the bulges and disks in our galaxies.   
We present B-, R- and I-band surface photometry for a subsample of 21
galaxies from \citetalias{Noordermeer05}, with morphological type ranging
from S0 to Sab and absolute B-band magnitudes $-17 > M_{\mathrm B} >
-22$.  
We have decomposed the images into contributions of a flattened disk
(including rings, bars, etc.) and a spheroidal bulge with a genuine 3D
structure, using an interactive 2D decomposition technique. 
The radial distribution of stars in each component is then carefully
measured and luminosity profiles are constructed and fitted with a
general S\'ersic profile (bulge) and exponential disk. 

In addition to their use in the mass models described above, our data also
provide interesting information about the internal structure of bulges and
their relation with the surrounding disks. 
We present an analysis on the various correlations between different bulge-
and disk parameters and briefly discuss the implications for different bulge
formation mechanisms. 

The structure of this paper is as follows. We describe the sample selection in
section~\ref{sec:sample}.  
The observations and data reduction steps are described in
section~\ref{sec:obs_and_redux}.    
In section~\ref{sec:isofits} we discuss the photometric analysis of the
galaxies, including the derivation of radial surface brightness profiles,
total magnitudes, diameters, etc. 
In section~\ref{sec:comparison}, the consistency of our photometric results is
checked by comparing different observations of the same galaxies. 
In section~\ref{sec:BD} we present the procedure that was developed to
decompose our images into bulge and disk contributions.  
We discuss the implications of our results in section~\ref{sec:discussion}. 
Finally, the individual galaxies in our sample are discussed in
section~\ref{sec:notes}, and a brief summary of our study and the main
conclusions are presented in section~\ref{sec:conclusions}.
For clarity, all long tables have been placed at the end of the
paper, in appendix~\ref{app:tables}. 
In appendix~\ref{app:atlas}, we show the images and photometric profiles for
all galaxies and the results from the bulge-disk decompositions.

\section{Sample selection}
\label{sec:sample}
The galaxies studied in this paper form a subset of the 68 galaxies with \HI\
observations from \citetalias{Noordermeer05}, which
themselves were selected from the WHISP survey \citep[Westerbork Survey of
\HI\ in spiral and irregular galaxies;][]{Kamphuis96,VanderHulst01}. 
The WHISP sample consists of galaxies on the northern hemisphere ($\delta >
20\deg$), with optical diameter and \HI\ line flux limits of $D_{25} >
1\arcmin$ and $f > 20 \, {\mathrm {mJy}}$ respectively.
The sample of \citetalias{Noordermeer05} consisted of all galaxies in the
WHISP survey with morphological type between S0 and Sab.
A more detailed description of the parent sample and the properties of
the selected galaxies is given in \citetalias{Noordermeer05}.

Given the context of the present paper as part of our larger study of rotation
curves and dark matter in early-type disk galaxies, the selection of galaxies
from the parent sample in \citetalias{Noordermeer05} was mainly based on the
requirement that good rotation curves could be derived. 
In practice, the following, somewhat subjective criteria were used: {\em 1)}
the \HI\ velocity field must be well resolved ($>$~5 -- 10 beams across) and
defined over significant parts of the disks (i.e.\ not confined to small
`patches'); {\em 2}) the gas must be moving in regular circular orbits around
the centre of the galaxy. Strongly interacting galaxies, strongly barred
galaxies, or galaxies with otherwise distorted kinematics are excluded; {\em
  3}) the inclination angle must be well constrained and preferably lie between
40\deg and 80\deg. 

Very few galaxies from \citetalias{Noordermeer05} satisfy all these
conditions and a strict application of these criteria would lead to a very
small sample.   
For the present paper, we have therefore relaxed the selection criteria and
included a number of galaxies with e.g.\ bars, mild kinematical
distortions or a more face-on orientation.    

Although the sample selection criteria described above may, at first sight,
appear somewhat irrelevant for a study of bulge and disk properties, we
believe that our galaxies form a representative sample of early-type disk
galaxies.  
In \citetalias{Noordermeer05}, we showed that the galaxies in the parent
sample resemble very much `average' early-type disk galaxies. 
The additional selection for the current paper, based on the kinematical
properties described above, is unlikely to introduce a bias towards any
particular bulge or disk properties. 

In total, the current sample consists of 21 galaxies; a few basic
characteristics of the members are given in table~\ref{table:sample}.  
The distances in the table were taken from \citetalias{Noordermeer05} and
were based on the systemic velocities derived there, corrected for
Virgo-centric inflow, and a Hubble constant of $75 \, {\mathrm {km \, s^{-1}
\, Mpc^{-1}}}$. 
The majority of the galaxies in our sample are luminous (16 systems have $M_B
< -20)$. 
Of the 21 galaxies, 16 are classed as spirals (type Sa or Sab) and 3 as
lenticulars (S0- -- S0+); the remaining two fall in the transitional S0/a
class. 
\addtocounter{footnote}{-1}

\begin{table}
 \centering
  \caption[Optical surface photometry sample: basic data]
  {Sample galaxies: basic data. (1)~UGC number; (2)~alternative name; 
   (3)~morphological type (taken from NED\footnotemark);
   (4)~absolute B-band magnitude; (5)~distance
   \citepalias[from][]{Noordermeer05}. \label{table:sample}}

   \begin{tabular}{rllrr}
    \hline
    \multicolumn{1}{c}{UGC} & \multicolumn{1}{c}{alternative} & 
    \multicolumn{1}{c}{Type} & \multicolumn{1}{c}{M$_{\mathrm B}$} & 
    \multicolumn{1}{c}{D} \\   
     
    & \multicolumn{1}{c}{name} & & \multicolumn{1}{c}{mag} &
    \multicolumn{1}{c}{Mpc} \\
    
    \multicolumn{1}{c}{(1)} & \multicolumn{1}{c}{(2)} &
    \multicolumn{1}{c}{(3)} & \multicolumn{1}{c}{(4)} &
    \multicolumn{1}{c}{(5)} \\ 
    \hline 
    
    89    & NGC 23   & SB(s)a       & -21.45 & 62.1 \\ 
    94    & NGC 26   & SA(rs)ab     & -20.32 & 62.6 \\ 
    624   & NGC 338  & Sab          & -20.83 & 65.1 \\
    1541  & NGC 797  & SAB(s)a      & -21.12 & 77.0 \\
    2487  & NGC 1167 & SA0-         & -21.88 & 67.4 \\ 
    2916  & --       & Sab          & -21.05 & 63.5 \\
    2953  & IC 356   & SA(s)ab pec  & -21.22 & 15.1 \\
    3205  & --       & Sab          & -20.89 & 48.7 \\
    3546  & NGC 2273 & SB(r)a       & -20.02 & 27.3 \\ 
    3580  & --       & SA(s)a pec:  & -18.31 & 19.2 \\
    3993  & --       & S0?          & -20.19 & 61.9 \\
    4458  & NGC 2599 & SAa          & -21.38 & 64.2 \\ 
    5253  & NGC 2985 & (R')SA(rs)ab & -20.93 & 21.1 \\
    6786  & NGC 3900 & SA(r)0+      & -19.93$^\dagger$\hspace{-0.12cm} & 25.9 \\
    6787  & NGC 3898 & SA(s)ab      & -20.00 & 18.9 \\
    8699  & NGC 5289 & (R)SABab:    & -19.48 & 36.7 \\ 
    9133  & NGC 5533 & SA(rs)ab     & -21.22 & 54.3 \\
    11670 & NGC 7013 & SA(r)0/a     & -19.20 & 12.7 \\
    11852 & --       & SBa?         & -20.44 & 80.0 \\
    11914 & NGC 7217 & (R)SA(r)ab   & -20.27 & 14.9 \\
    12043 & NGC 7286 & S0/a         & -17.53 & 15.4 \\ 
    \hline 
    \multicolumn{5}{l}{$^\dagger$ No B-band data available in this
     study; M$_{\mathrm B}$ taken} \\
    \multicolumn{5}{l}{\hspace{0.24cm}from LEDA\footnotemark.} \\     
   \end{tabular}
\end{table}  

\section{Observations and data reduction}
\label{sec:obs_and_redux}
We have obtained images of all our galaxies in Harris R- and B-bands,
with a few additional observations in Harris I-band; for UGC~6786,
only R-band data were available. 
\addtocounter{footnote}{-1}
\footnotetext{NED: The NASA/IPAC Extragalactic Database
  (http://nedwww.ipac.caltech.edu), is operated by the Jet propulsion
  Laboratory, California Insitute of Technology, under contract with the
  National Aeronautics and Space Administration.} 
\addtocounter{footnote}{+1}
\footnotetext{LEDA: The Lyon Extragalactic Database,
  http://leda.univ-lyon1.fr/}  
The bulk of the observations described here were performed during 4
observing runs in 2000 and 2001 with the 1.0m Jacobus Kapteyn Telescope (JKT)
at the Observatorio Roque de los Muchachos on La Palma\footnote{The 
  Jacobus Kapteyn Telescope and Isaac Newton Telescope are operated on the
  island of La Palma by the Isaac Newton Group in the Spanish Observatorio del
  Roque de los Muchachos of the Instituto de Astrofisica de Canarias. Part of
  the data described in this study was collected via the service observation
  scheme of the Isaac Newton Group.}.  
A few observations were performed at the JKT in service mode in 2002.  
In all cases we used the JAG-CCD camera which was equipped with a 2048\keer 
2048 SITe chip with a pixel size of 0.33\arcsec\ and a total unvignetted
field-of-view of approximately 10\arcmin\keer 10\arcmin. 

\addtocounter{footnote}{-1}
A small number of observations were done using the Wide Field Camera on the
2.5m Isaac Newton Telescope (INT) on La Palma\footnotemark.  
It was equipped with four 4096\keer 2048 EEV CCD's with 0.33\arcsec\
pixels, offering a total field of view of 34\keer 34\arcmin. 

Finally, a few images were taken with the 8K Imaging Camera at the
2.4m MDM Hiltner telescope at Kitt Peak, Arizona. 
It contains a mosaic of eight 4096\keer 2048 SITe CCD's with 0.18\arcsec\
pixels and a total field of view of 24\keer 24\arcmin.   

Typical exposure times for all observations on the JKT were 30 -- 50 minutes
per band, usually broken up into at least 2 shorter exposures to enable
cosmic ray rejection.  
Sometimes, more but shorter exposures were taken to prevent the bright cores
of the images from saturating on the chip.  
For the observations at the larger INT and MDM telescopes, we used
correspondingly shorter exposures to achieve roughly the same sensitivity as
with our exposures on the JKT.

During some nights the observing conditions were non-photometric due to thin
cirrus-clouds, and in some other nights the seeing was very poor. 
Useful images could still be taken on those nights, but in the former case the
absolute photometric calibration was not reliable, whereas in the latter case
the central light concentrations were poorly resolved. 
Those problems were fixed by re-observing the galaxies with short exposure
times later.  
A few galaxies were observed in good conditions several times. 
Those observations are used in section~\ref{sec:comparison} to check the
consistency of our observations.  
In table~\ref{table:observations} we list the observations of all
galaxies.\\

Data reduction was performed within the IRAF environment\footnote{IRAF is
  distributed by the National Optical Astronomy Observatories, which are
  operated by the Association of Universities for Research in Astronomy, Inc.,
  under cooperative agreement with the National Science Foundation.}. 
Standard procedures were followed to subtract readout bias and apply flatfield
corrections. 
Separate exposures for each galaxy were aligned using a number of bright
stars.  
They were then combined, using a simple rejection criterion to remove cosmic
ray events and cosmetic defects on the chip.  

Special care was taken in the subtraction of the background light. 
In the first three runs on the JKT, the camera suffered from
light-leaks which caused faint residual gradients in the background
after flatfielding, especially in the R-band images. 
In almost all cases, the residuals could be removed by fitting a
first- or second-order 2D polynomial to the background and subtracting it from
the image.   
In a few cases we used polynomials up to order 5.  
In all cases, extreme care was taken to exclude the image of the
galaxy from the fitting region, so as not to subtract any light from
the galaxy itself.  

In all other observations, including the B-band observations from the
JKT runs with light-leaks, the flatfielding worked very well and
residuals were small.   
For these images, the background was removed by fitting a first-order
polynomial to the emission-free regions.

As an extra check that the light-leaks did not affect our results, we 
re-observed UGC~2953 and 3546 in the fourth observing run on the 
JKT, when the light-leaks were eliminated. 
Comparison of the resulting photometric profiles from the images with
and without light-leaks shows that the gradients were adequately removed and
that the resulting profiles are consistent within the errors (see
section~\ref{sec:comparison}). 

Throughout the nights, standard star fields from \citet{Landolt92}
were observed at intervals of 1 -- 2 hours. 
Using the instrumental fluxes of the stars and the magnitudes given by 
Landolt, we fitted the magnitude zero-points and extinction
coefficients, such that 
\begin{equation}
m_{\mathrm{Landolt}} = m_{\mathrm{obs}} + m_0 + e \cdot {\cal{A}},
\label{eq:photcal} 
\end{equation}
where $m_{\mathrm{Landolt}}$ is the stellar magnitude given by Landolt,
$m_{\mathrm{obs}}$ is our instrumental magnitude, $m_0$ is the magnitude
zero-point, $e$ is the extinction coefficient and ${\cal{A}}$ is the airmass.  
The coefficients were determined for each night and each colour separately.  
The errors $\Delta m_0$ and $\Delta e$ on the fitted coefficients were used to
derive the photometric error for each observation: 
\begin{equation}
 \sigma_{\mathrm {phot}} = \sqrt{ \Delta m_0^2 + (\Delta e \cdot
 {\cal{A}})^{2} }. 
 \label{eq:photerror}
\end{equation}
The resulting errors for all observations are given in
table~\ref{table:observations}.  
The average photometric errors for our B-, R- and I-band observations
are respectively 0.13, 0.11 and 0.11~mag, and respectively 90, 76 and
83\% have \mbox{$\sigma_{\mathrm {phot}} < 0.2\, {\mathrm {mag}}$}. 

Coordinate systems were added to the images by registering a number of bright
stars with known coordinates from the Digitized Sky Survey (DSS). 
Typical rms errors were of the order of 0.15\arcsec\ for the JKT images and
0.2\arcsec for the larger INT and MDM images.
Note, however, that the uncertainty in the coordinate system of the DSS is
about 0.6\arcsec\ \citep{Russell90, Veron-Cetty96}. 
Thus, the coordinate systems of our images are effectively limited by the 
uncertainties in the DSS and have an accuracy of no better than 0.6\arcsec.   

Finally, the images were cleaned from foreground stars, background galaxies
and any possible artifacts not removed in the previous steps. 
For the foreground stars, we fitted a 2D analytical PSF to $\approx \!
20$ bright, unsaturated stars on the image. 
Subsequently, all point sources in the image were detected, and the
flux in each was determined.  
The PSF was then scaled to the measured flux and model stellar
`images' were subtracted from the image. 
This proved to work in a satisfactory manner only for relatively faint
stars.  
For bright stars, the analytical PSF often did not represent the
stellar images sufficiently well, and significant residuals remained 
present in the image after subtraction. 
Saturated stars, background galaxies and other artifacts in the images
could, naturally, not be removed with this method either.  
All these were removed by hand, masking out the affected regions from
the image.  
Special care was taken to remove faint haloes of bright stars.  
The same masks were used for all images of a galaxy in different
bands, to ensure that the measured magnitudes and surface brightnesses
refer to the same regions. 

\section{Photometric properties}
\label{sec:isofits}
Photometry was performed at concentric elliptical annuli, using the ELLIPS
task from the STSDAS ISOPHOTE package \citep{Jedrzejewski87}. 
The ellipses were defined at logarithmic intervals in semi major axis,
starting at $\approx \! 0.3\arcsec$ and increasing the radius of each 
successive ellipse by a factor 1.1.

\subsection{Isophote orientation parameters}
\label{subsec:isoorient}
The orientation parameters of the ellipses were determined from the
R-band images in two steps. 

First, ellipses were fit to the image with the position of the centre,
position angle and ellipticity of each ellipse left as free fitting
parameters.   
From these fits, the central positions of the galaxies were
determined. 
The bulges of most galaxies create a clear central peak in the image
and the centres of the inner ellipses converge usually to a narrowly
defined value; often the variation between the inner ellipses is
smaller than 0.1\arcsec. 
In these cases, the error in the absolute central position is dominated by the  
uncertainty in the coordinate system (see section~\ref{sec:obs_and_redux}).  
In a small number of cases dust obscures the very centre of the image,
and the light distribution is more irregular. 
Since in those cases the brightest pixels lie at the position of least
obscuration, rather than at the true centre of the underlying light
distribution, the inner isophotes can not be used to determine the
centre and we rather used the ones at intermediate radii where the
effect of dust is either smaller or averages out over the ellipse.
The central positions of those galaxies are, however, necessarily less
accurately determined than for the ones where we can see into the
centre directly. 
The fitted centres are listed in columns (2) and (3) in
table~\ref{table:isoresults}; the uncertainty and its
dominant source are given in columns (4) and (5) respectively. 

In the second step, we did fits with the centre of each ellipse fixed
at the position just determined, but with ellipticity and position
angle still free.  
The fitted orientation parameters are shown by the symbols in the top
right hand panels in the figures in appendix~\ref{app:atlas}.  
In the outer part of the galaxies, outside the region where bulges and 
bars complicate the picture, they usually converge to more or less
constant values (see UGC~3205, 3580 and 6787 for nice examples). 
These values were then assumed to represent the true position angle
and inclination of the galaxy. 
In some cases, however, the position angle and/or ellipticity never
converge, and either keep varying till the last point (e.g.\ UGC~89, 1541 or
4458) or show several plateaus (e.g.\ UGC~9133).
In these cases, we visually compared the fitted isophotes to the shape 
of the galactic disks and then estimated the position angle and 
ellipticity by eye.
Note that large variations in the orientation parameters in the very
outer points (e.g.\ UGC~3993, 4458, 8699) can usually be attributed to
asymmetries in the light distribution of the galaxy or to imperfect
flatfielding; these variations were generally not considered real.   
The adopted values for the position angle and ellipticity are indicated
with the dashed lines in the figures in appendix~\ref{app:atlas} and listed in
columns (6) and (7) in  table~\ref{table:isoresults}.   

From the ellipticity we determined the inclination of the galaxies
using the standard formula
\begin{equation}
   \cos^2 i = \frac{(1-\epsilon)^2 - q_0^2}{1-q_0^2},
\label{eq:inclination}
\end{equation}
with $q_0$ the intrinsic flattening of the disk, for which we assumed a value
of 0.2 \citep[cf.][]{DeGrijs98}.  
The derived values are given in column (8) in table~\ref{table:isoresults}.  
Note that they are estimates only; if the disks have different intrinsic
flattening, or if they are intrinsically elongated rather than axisymmetric, 
than the true inclination may deviate significantly from these values.  

\subsection{Surface photometry}
\label{subsec:surfphot}
Once the orientation parameters were determined, we fixed them for all 
ellipses and simply measured the average intensities at each
radius. The same ellipses were used for the images in different colour
bands. 

\begin{figure}
 \centerline{\psfig{figure=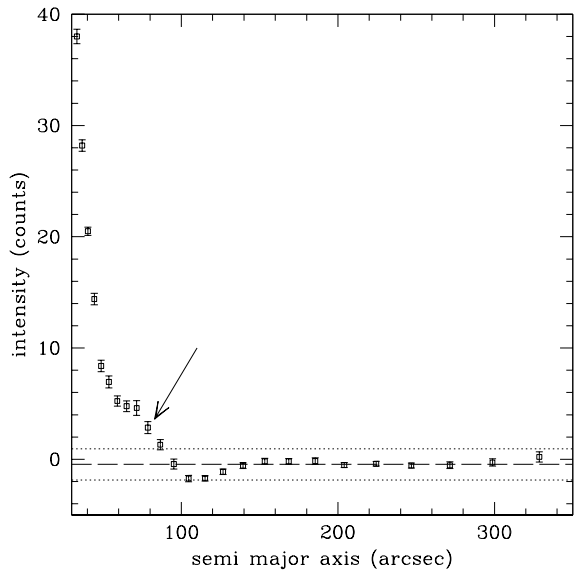,width=7.5cm}}
 \caption{Example of the determination of the exact background level for one
  of our images. The squares give the average intensities on each ellipse; the
  errorbars give the measurement error. The long-dashed line gives the average
  intensity of the outer ellipses, which defines the zero-point $I_{\mathrm
  {bg}}$ of the image. In this case, $I_{\mathrm {bg}} =  -0.46$, indicating
  that we have slightly over-subtracted the background. The dotted lines give
  the assumed error in the zero-point. The arrow indicates the outer point
  with significant flux from the galaxy. See text for
  details. \label{fig:background}} 
\end{figure}

Even though the background subtraction described in
section~\ref{sec:obs_and_redux} usually worked quite well, some small
residuals always remained.  
To determine the exact zero-points in the images, as well as their
uncertainties, we let the radii of the ellipses expand until they covered the 
entire images.    
Thus, intensities were measured not only on the galaxies, but also in annuli
centered on, but much larger than, the galaxies.  
A typical result is shown in figure~\ref{fig:background}. 

In this case, it is clear that no light from the galaxy is detected
beyond a radius of about 90\arcsec, so the ellipses outside this
radius probe the sky background.  
The average intensity on the outer ellipses defines the zero-point
$I_{\mathrm {bg}}$ of the image and is subsequently subtracted from
the intensities in the galaxy.  

The variation in the intensities on the outer ellipses measures the
uncertainty in the zero-point. 
Note, however, that these outer ellipses cover large parts of the image, such
that small scale irregularities in the background are averaged out. 
To account for irregularities at smaller scales as well, we adopted for the
zero-point uncertainty $\sigma_{\mathrm {bg}}$ {\it twice} the rms variation
in the intensities of the outer ellipses.   
In table~\ref{table:observations} we give the magnitude level $\mu_{3\sigma}$
corresponding to $3 \sigma_{\mathrm {bg}}$ above the sky; the average values
are 26.95, 26.07 and 24.26\magasas\ in the B-, R- and I-band images
respectively. 

The errors on the measured intensities in the galaxy are now given by
the quadratic sum of the measurement error $\Delta I_m(r)$ and the
zero-point error $\sigma_{\mathrm {bg}}$:
\begin{equation}
 \sigma_I(r) = \sqrt{\Delta I_m(r)^2 + \sigma_{\mathrm {bg}}^2}
\end{equation}

Finally, the intensities and corresponding errors were calibrated and 
converted to surface magnitudes using the transformations determined 
in section~\ref{sec:obs_and_redux}.  
Only points with intensities larger than $2\sigma_{\mathrm {bg}}$ were
considered. 
The resulting photometric profiles are shown in the top middle panels on the
first row of the figures in appendix~\ref{app:atlas}.  
In the bottom middle panels on the first row of the figures, the colour
profiles are shown.   
The errors on the points are given by the quadratic sum of the errors on the
individual bands. 
Points are only shown when the total error is smaller than 0.5 mag.  

\subsection{Deviations from axisymmetry}
\label{subsec:asymmetries}
In addition to the average intensity on each ellipse, we also measured the
higher-order harmonic components to study deviations from perfect
axisymmetry. 
The intensity distribution along each ellipse was decomposed into a Fourier
series of the following form: 
\begin{equation}
 I(\alpha) = I_0 + I_1 \cos(\alpha + \phi_1) + I_2 \cos\left(2(\alpha +
 \phi_2)\right) + \ldots,
\end{equation}
where $\alpha$ is the azimuthal angle along the ellipse, $I_0$ is the average
intensity on the ellipse and $I_1$ and $I_2$ measure the strength of the m=1
and m=2 Fourier components respectively. 
Thus, a high value for $I_1/I_0$ indicates a strongly lopsided light
distribution, whereas a high value for $I_2/I_0$ reveals the presence of a bar
or oval distortion.

A statistically significant measurement of these higher order terms in the
light distribution requires a higher signal-to-noise ratio than is necessary
for the zeroth order term. 
The m=1 and m=2 terms were therefore only measured on ellipses for which the
average intensity was respectively $2\!\sqrt2$ and 4 times $\sigma_{\mathrm
{bg}}$.   

The relative strengths and the phases of the m=1 and m=2 components are
shown in the bottom 4 panels on the right side of the first row of the figures
in appendix~\ref{app:atlas}.

\subsection{Isophotal diameters, effective radii and total magnitudes}
\label{subsec:totalmag+diam}
For each galaxy and each band, several diameters were derived. $D_{25}$ and
$D_{26.5}$ are isophotal diameters, determined at $\mu = 25$ and
$26.5\magasas$ respectively. $R_{20}$, $R_{50}$ and $R_{80}$ are effective
radii, containing respectively 20, 50 and 80\% of the light. 
The diameters and effective radii are given in table~\ref{table:isoresults}.  

Note that these diameters and effective radii are derived from the
photometric profiles as measured on the sky. 
However, the surface brightnesses of the disks of the galaxies are dimmed by
Galactic foreground extinction. 
On the other hand, the observed surface brightnesses are higher than the true
values due to the fact that we observe the galaxies under a non-zero
inclination angle. 
The raw isophotal diameters derived above do therefore not correspond to the 
same physical surface brightness in the disks of these galaxies. 
In table~\ref{table:isoresults} we also give the isophotal diameters
$D^{c}_{25}$ and $D^{c}_{26.5}$ that correspond to face-on, extinction
corrected levels of $\mu = 25$ and $26.5\magasas$. 
The correction for inclination was performed assuming that the galactic disks
are optically thin; no corrections were made for internal extinction caused by
dust in the galactic disks themselves.  
The corrections for Galactic foreground extinction were performed using the
values from \citet{Schlegel98}.   

Total magnitudes cannot be derived from the images directly, as large parts of
the images are sometimes masked to remove foreground stars, background
galaxies, etc. (section~\ref{sec:obs_and_redux}), such that the masked
images contain flux from parts of the galaxies only.  
To correct for this effect, model images were created, based on the
photometric profiles derived in section~\ref{subsec:surfphot}, interpolating
over the masked regions. 
From these model images, two apparent magnitudes were derived. 
$m_{\mathrm{lim}}$ is the total magnitude within the last measured point in
the photometric profile; $m_{25}$ is the apparent magnitude within the
25th\magasas\ isophotal diameter $D_{25}$.  
The errors in the total magnitudes are usually dominated by the uncertainty
$\sigma_{\mathrm {phot}}$ in the photometric calibrations
(equation~\ref{eq:photerror}, column (7) in table~\ref{table:observations}),
with small contributions from the zero-point uncertainty $\sigma_{\mathrm
{bg}}$ and Poisson errors.  

The corresponding absolute magnitudes $M_{\mathrm{lim}}$ and $M_{25}$ were
determined using the distances as given in table~\ref{table:sample} and were
corrected for Galactic foreground extinction using the values from
\citet{Schlegel98}.

\section{Internal comparison of photometric profiles}
\label{sec:comparison}
Several galaxies were observed during different nights and with different
telescopes (see table~\ref{table:observations}).  
For each of these cases, only the best data are shown in
appendix~\ref{app:atlas} and listed in tables~\ref{table:isoresults} and
\ref{table:photresults}.  
The data from the other observations have been used to assess the quality of
our data and the reliability of the derived errors. 
These observations were reduced independently, following the same steps as for
the main observations.  
The same masks were used, however, for all images
(section~\ref{sec:obs_and_redux}), to make sure that measured surface
brightnesses correspond to the same regions in all images.  
Similarly, the same position angles and ellipticities were used to derive the
photometric profiles.   

\begin{figure*}
  \centerline{\psfig{figure=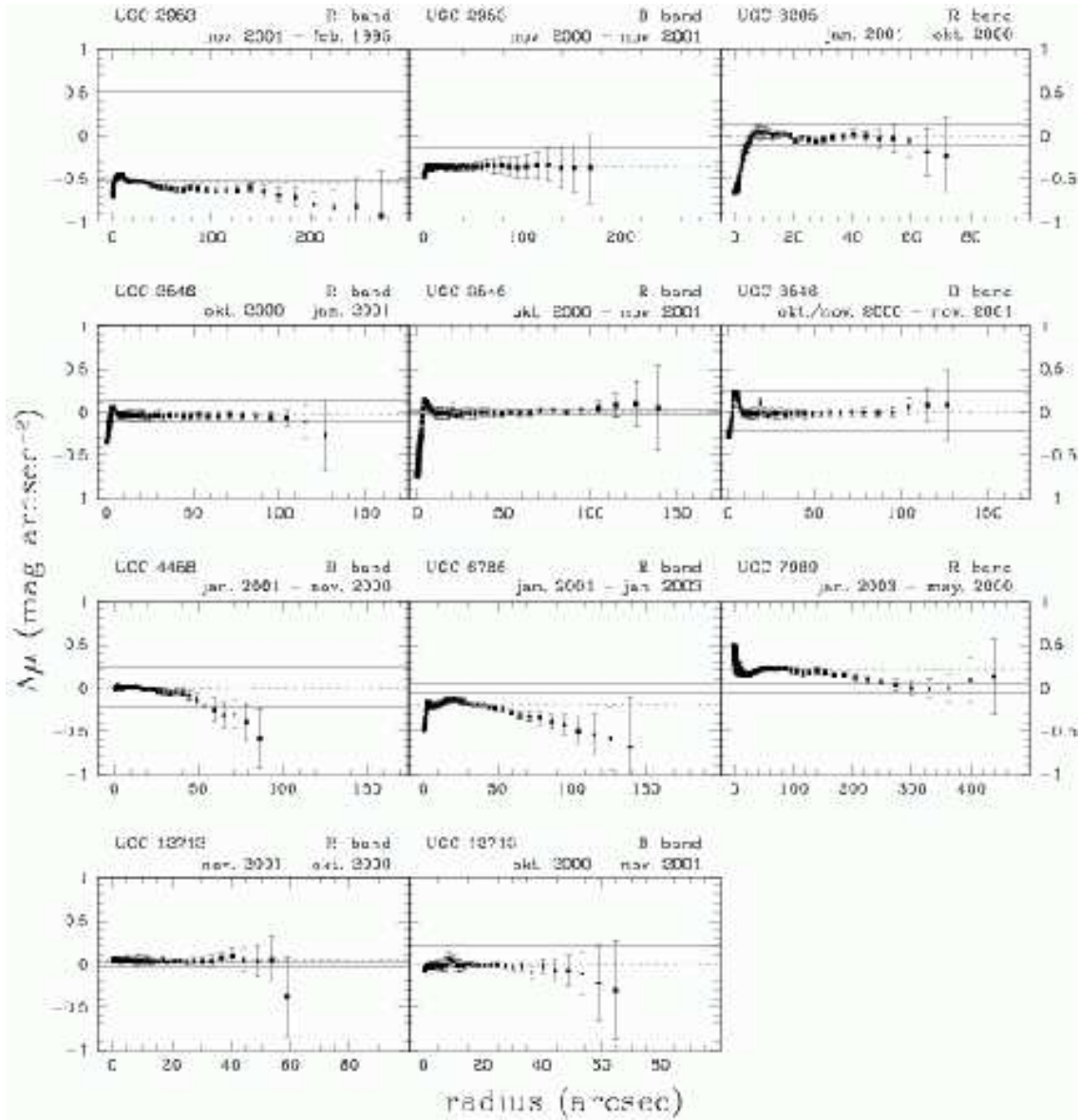,width=15cm}}
  \caption{Internal comparison of the photometric profiles. Data points
  show the difference between photometric profiles from multiple 
  observations of the same galaxy. The observations which are compared
  are indicated above each panel; see table~\ref{table:observations} for
  details. Errorbars give the combined errors on each point. Dashed lines give
  the weighted average of the points which are not affected by seeing
  differences. Solid lines give the combined $1\sigma$ photometric errors. 
  \label{fig:intcomp}} 
\end{figure*}

In figure~\ref{fig:intcomp}, we compare the photometric profiles for all the
galaxies with multiple observations.  
In most cases, the independently derived profiles agree very well, but
there are also some cases with significant differences.

Errors in the photometric calibrations lead to constant offsets in a
profile but leave its shape unchanged. 
Indeed, systematic offsets between 2 profiles are seen for some
galaxies in figure~\ref{fig:intcomp} (the dotted lines give the
weighted average of the data points), but in most cases, these offsets
fall within the combined $1\sigma$ photometric errors (solid lines). 
In a few cases the offsets are larger than to be expected (notably
UGC~2953, 6786 and 7989\footnote{The analysis in this section includes two
  galaxies, UGC~7989 and 12713, which were observed during the same runs as
  the other galaxies in our sample, but which are not included in the
  remainder of this study. They are only shown here as additional
  comparison material.}). 
In the case of UGC~2953, the differences between the independent
observations can be explained as the result of non-photometric
conditions during the nights of the comparison observations. 
In tables~\ref{table:observations}, \ref{table:photresults} and
\ref{table:BDresults}, we have marked all observations that were done under 
non-photometric conditions. 
Note that non-photometric conditions do not affect the shapes of the
photometric profiles. 
Scale lengths, effective radii, bulge-disk ratios, etc.\ can still be
accurately measured on such nights. 
Furthermore, in many cases, galaxies that were observed under non-photometric
conditions were later re-observed with short exposure times; these short
exposures were then used to calibrate the deep images taken before and
accurate photometry could still be achieved. 

The offsets for UGC~6786 and 7989 are probably caused by differences between
the filter transmission curves used for the observations on the JKT and the
MDM telescope. 
Note that the offsets between the JKT and MDM profiles are equally large
(0.20\magasas) for both UGC~6786 and 7989; this strengthens the interpretation
of the offsets as a result of filter differences.   

Apart from the systematic offsets between the profiles, random point-to-point
variations are generally small, except for the inner few arcseconds where
differences in seeing between the observations lead to artificially large
discrepancies.  
In some cases, errors in the sky-level determination lead to differences in
the outer points of the compared profiles, but they are generally within the
errors derived as described in section~\ref{subsec:surfphot}. 
The largest sky-subtraction errors are seen in UGC~4458, where the outer
points of the two compared profiles disagree at a level of 2 times the
combined $1\sigma$ errors.  
Given the number of profiles compared here, one case of $2\sigma$ disagreement
is to be expected.  
The only cases where we observe significant differences between the profiles
that cannot be attributed to seeing effects or uncertainties in the sky-level
are the profiles which were observed with different telescopes (R-band
observations of UGC~2953, 6786 and 7989); as above, these differences can
probably be attributed to the different filters which were used for the
observations under comparison here.   

In conclusion, for galaxies that were observed under photometric conditions,
the photometric errors listed in table~\ref{table:observations} seem to be
reliable; for non-photometric observations the errors are lower limits. 
The errors on individual data points in our photometric profiles are also
realistic and account well for the uncertainties in the determination of the
sky level.   
Photometric profiles which were derived from images observed with the INT or
MDM telescope show small deviations ($\sim 0.1-0.2 \magasas$) compared to the
profiles derived from the JKT images.  
The deviations manifest both as systematic offsets between the profiles, as
well as point-to-point variations within the profiles.   
These differences are probably caused by differences in the filters used for
the observations.

\section{Bulge-disk decompositions}
\label{sec:BD}
Many methods exist to decompose the light of spiral galaxies into
contributions from bulges and disks.  
Traditionally, the decomposition is performed on the photometric
profiles directly, fitting them with the sum of an exponential disk
and a certain profile for the bulge (usually either an $r^{1/4}$, 
exponential or general S\'ersic profile).  
As this is a 1D procedure, it is quick and can therefore be used to
study large samples of galaxies in short timespans
\citep[e.g.][]{Baggett98,Graham01, MacArthur03}.
However, photometric profiles suffer from projection effects. 
The observed intensity at each point in a galaxy is a superposition of 
light from the bulge and disk.
Because they have different intrinsic shapes, the contributions from
both components come from different radii when the galaxy is observed
under a non-zero inclination angle, and the average intensity along
a given isophote is generally not directly related to the true mean
brightness at that radius.   
Thus, deriving bulge and disk parameters from azimuthally averaged
photometric profiles will lead to systematic errors.
A proper treatment of the projection effects requires a full 2D
decomposition technique \citep[e.g.][]{Byun95,DeJong04,
Laurikainen05}.   

\begin{figure}  
 \centerline{\psfig{figure=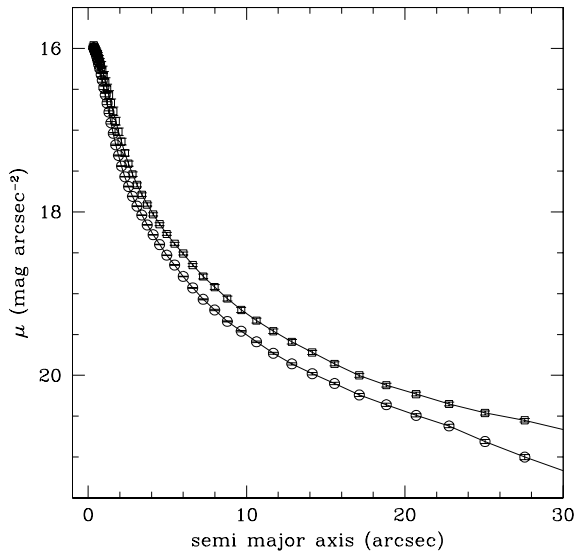,width=7.5cm}}
 \caption{The influence of projection effects on the central R-band
  photometric profile of UGC~6786. The profile shown with squares is
  derived using the standard method, i.e.\ the intensities are measured 
  along ellipses with ellipticity fixed at the value of the outer
  regions. The circles denote the profile when measured along ellipses
  with an ellipticity which better represents the isophotes of the
  bulge. 
  \label{fig:projection}}
\end{figure}

The projection effects described above are particularly severe in the
type of galaxies studied here, which have often large and luminous
bulges.  
As an illustration, we show in figure~\ref{fig:projection}
two R-band photometric profiles of the central part of the
bulge-dominated galaxy UGC~6786. 
The profile shown with squares is derived following our standard
procedures of section~\ref{subsec:surfphot}, that is, it 
shows the average intensities measured along ellipses with the
ellipticity fixed at the value of the outer regions.  
The circles show the intensities when measured along ellipses with a much
lower ellipticity, which is more representative of the isophotes in the
central, bulge-dominated region. 
Clearly, the bulge of UGC~6786 is more centrally concentrated than is
shown by our standard photometric profile; any structural parameters of the
bulges derived directly from our photometric profiles will be severely
affected by systematic errors.   

Another problem with bulge-disk decompositions is the degeneracy that
exists between the different parameters \citep{MacArthur03,DeJong04}.
The data can often be fitted by different combinations of parameters, so that
it can be difficult to reach unambiguous conclusions about the true values. 

To overcome the difficulties mentioned above, we have developed a hands-on,
interactive procedure that uses the full 2D information from the images to
separate them into bulge and disk components.  
Our method bears some similarities with the one used by \citet{Palunas00} and
can be summarised as follows.  
We model the bulges as flattened axisymmetric spheroids, with intrinsic axis
ratios $q_b$ and general S\'ersic photometric profiles characterised by
effective radius and magnitude $r_e$ and $\mu_e$ and shape parameter $n$
\citep{Sersic68}.   
The bulge parameters are determined from images where an initial
estimate for the disks has been subtracted. 
Based on the fitted parameters, model images of the bulges are created 
and subtracted from the original images. 
All light in these image is then assumed to originate from the disks, 
and their photometric profiles are used to derive the disk light
distributions.   

A more detailed description of the separate steps is given in the
following subsections.

\subsection{Bulge parameters}
\label{subsec:bulges}
The first step in our procedure consists of deriving initial estimates
for the bulge and disk parameters from an analytic fit to the
photometric profile.  
As discussed above, the central, bulge-dominated parts of the
photometric profiles suffer from projection effects.
In the outer parts, however, the profile is usually dominated by the
flat disk, and projection effects are not an issue.
Thus, the initial estimates for the disk are generally sufficiently
reliable for the next step.

Based on the initial estimates, an initial model image of the disk component
is created by extrapolating the fitted disk profile inwards. 
This model image is then subtracted from the original image to obtain
an image with, to first order, bulge-light only. 

From the resulting image, the bulge-parameters can be obtained. 
We first determine the flattening of the bulge by fitting ellipses to
the R-band bulge image with ellipticity and position angle as free
parameters.  
In most cases, the position angle of the bulge isophotes is close to
the value derived for the outer disk, indicating that the bulge is a
flattened spheroid with the plane of symmetry coinciding with that of
the disk. 
In a few cases, the position angle in the centre differs from the
outer values, indicating the presence of a nuclear bar or other
triaxial structure. 
The presence of these non-axisymmetric structures is ignored in the
following; we model all bulges as oblate spheroids that are aligned
with the outer disk and we simply average out any non-axisymmetric
structures along the isophotes in the next step.
Because the bulges are more vertically extended than the surrounding 
disks, their ellipticities are generally lower than those found for
the outer disks. 
From the average ellipticity $\epsilon_b$ of the bulge isophotes, the
intrinsic flattening of the bulge can be determined by rewriting
equation~\ref{eq:inclination} as 
\begin{equation}
  q_b^2 = \frac{(1-\epsilon_b)^2 - \cos^2 i}{1 - \cos^2 i}.
  \label{eq:bulgeflattening}
\end{equation}
Here $q_b$ is the intrinsic axis ratio of the bulge and $i$ is the
inclination of the system, as derived in
section~\ref{subsec:isoorient}. 

We next derive photometric profiles for the bulge in all available colour
bands by measuring the average intensities on concentric ellipses with fixed
ellipticity $\epsilon_b$.  
Intensities are only measured inside the radius where the bulge and disk
intensities from the original image are equal.  
The resulting profiles are shown with the data points in the figures in
appendix~\ref{app:atlas}.  

To extrapolate the contribution of the bulge to larger radii, we
fitted the profiles with a general S\'ersic light profile
\citep{Sersic68}: 
\begin{equation}
 I_b(r) = I_{0,b} \exp \left[ - \left( \frac{r}{r_0} \right) ^{1/n} \right].
\label{eq:Sersic0}
\end{equation}
Here, $I_{0,b}$ is the central surface brightness and $r_0$ is the
characteristic radius. 
$n$ is a shape parameter, that describes the curvature of the profile
in a radius-magnitude plot. 
For $n=4$, equation~\ref{eq:Sersic0} reduces to the
well-known De Vaucouleurs profile \citep{DeVaucouleurs48},
whereas $n=1$ describes a simple exponential profile.
In the literature, equation~\ref{eq:Sersic0} is usually
written as 
\begin{equation}
 I_b(r) = I_e \exp \left[ - b_n \left\{ \left( \frac{r}{r_e} \right)
 ^{1/n} \! - 1 \right\} \right],
\label{eq:Sersiceff}
\end{equation}
with $I_e = e^{-b_n} I_{0,b}\,$ the intensity at the effective radius $r_e =
b_n^{\; n} \, r_0$, the radius which encompasses 50\% of the
light. 
$b_n$ is a scaling constant that is defined such that it satisfies
$\gamma(b_n, 2n) = \frac{1}{2} \Gamma(2n)$, with $\gamma$ and $\Gamma$ the 
incomplete and complete gamma functions respectively. 
It can be approximated by $b_n \approx 1.9992 n - 0.3271$ for $1 < n <
10$ \citep{Graham01}, \label{eq:bn} but in our fitting procedure, we
determined $b_n$ more accurately by numerically solving the equation above.   

In magnitudes, equation~\ref{eq:Sersiceff} is written as 
\begin{equation}
 \mu_b(r) = \mu_e + 1.0857 \, b_n \left\{ \left( \frac{r}{r_e} \right)
 ^{1/n} \! - 1 \right\}. 
\label{eq:Sersicmag}
\end{equation}
Before fitting the profile of equation~\ref{eq:Sersicmag} to
the observed data, it is corrected for seeing effects using the
convolutions from \citet{Graham01}. 
 
The seeing-convolved profile is then fitted to the data using a standard
non-linear least-squares algorithm. For the fits to the B- and I-band
profiles, we fixed the parameters $n$ and $r_e$ to the values found
for the R-band profile, leaving only the effective surface magnitude
$\mu_e$ free. 
Thus, we model the bulges as fully mixed systems without colour
gradients.   
Any remaining colour gradients after the subtraction of the bulge
influence are attributed to the disk.  
The fitted profiles are shown as the solid lines in the figures in 
appendix~\ref{app:atlas}.

The fitted bulge parameters $\epsilon_b$, $q_b$, $\mu_e$, $r_e$ and $n$
are given in table~\ref{table:BDresults}. 
The effective surface brightness was also corrected for Galactic
foreground extinction, using the values from \citet{Schlegel98}; the
corrected surface magnitude $\mu_e^c$ is listed in
table~\ref{table:BDresults} as well.
Finally, a total magnitude was derived by integrating
equation~\ref{eq:Sersic0} to infinity; the resulting values
$m_b$ and $M_b$ for the apparent and absolute magnitudes are also
listed in table~\ref{table:BDresults}. 
Model images for the bulge in the R-band, created using the parameters
found above, are shown in the bottom left panels in the figures in
appendix~\ref{app:atlas}.   

\subsection{Disk parameters}
\label{subsec:disks}
To measure the light distribution in the disk, the model images of the
bulge are subtracted from the original images in each available
colour. The resulting R-band images are shown in the bottom right
panels in the figures in appendix~\ref{app:atlas}.  

All remaining light in these images is then assumed to originate from
the disk component.  
This also includes residuals in the bulge-dominated parts, where our
fitted analytical profile (equation~\ref{eq:Sersiceff}) does
not exactly match the observed light distribution. 
Note, however, that the deviations of the observed bulge profile from
the analytic fit are almost always very small compared to the total
over-density of bulge-light over the inner disk (typically 0.05 -- 0.1 vs.\ 2
-- 4 magnitudes). 
The error caused by attributing the residuals to the disk, instead of
to the bulge, will be negligible when calculating the combined
dynamical impact of both components.

The radial distribution of the disk light is then determined by
measuring the average intensities on the ellipses defined in
section~\ref{subsec:isoorient}. 
The resulting photometric profiles for the disk, outside the
bulge-disk transition radius, are shown in the bottom right panels in
the figures in appendix~\ref{app:atlas}.  
In two cases, UGC~6786 and 6787, it was found, after the subtraction
of the bulge, that the ellipticity derived from the original image did 
not accurately represent the inclination of the disk.
In these two cases, the bulges are so dominant that, in the original
images, they influence the shape of the isophotes even in the outer
parts. 
The correct inclination angle of the disks could therefore only be
determined after the subtraction of the bulge, and for the final disk
profile, we used values different than the ones listed in
table~\ref{table:isoresults} (see individual notes in
section~\ref{sec:notes}). 
The corrected inclination angle was also used to recalculate the
flattening of the bulges for these galaxies
(equation~\ref{eq:bulgeflattening}). 

Finally, exponential profiles were fitted to the observed disk
profiles:
\begin{equation}
 I_d(r) = I_{0,d} \exp \left(- \frac{r}{h} \right),
\end{equation}
or in magnitudes:
\begin{equation}
 \mu_d(r) = \mu_{0,d} + 1.0857 \left( \frac{r}{h} \right),
\end{equation}
with $I_{0,d}$ and $\mu_{0,d}$ the central surface brightness, and $h$
the radial scale length. 
Only points outside the bulge-disk transition radius were used for the
fit. 
The resulting parameters are listed in
table~\ref{table:BDresults}; the corresponding profiles are
shown as the solid lines in the figures in
appendix~\ref{app:atlas}.
In table~\ref{table:BDresults}, we also list $\mu_{0,d}^c$,
the central surface brightness corrected for inclination and Galactic 
foreground extinction. 
The correction for inclination was performed assuming that the
galactic disks are optically thin; the corrections for Galactic
foreground extinction were performed using the values from
\citet{Schlegel98}.

\section{Discussion}
\label{sec:discussion}
\begin{figure*}
  \centerline{\psfig{figure=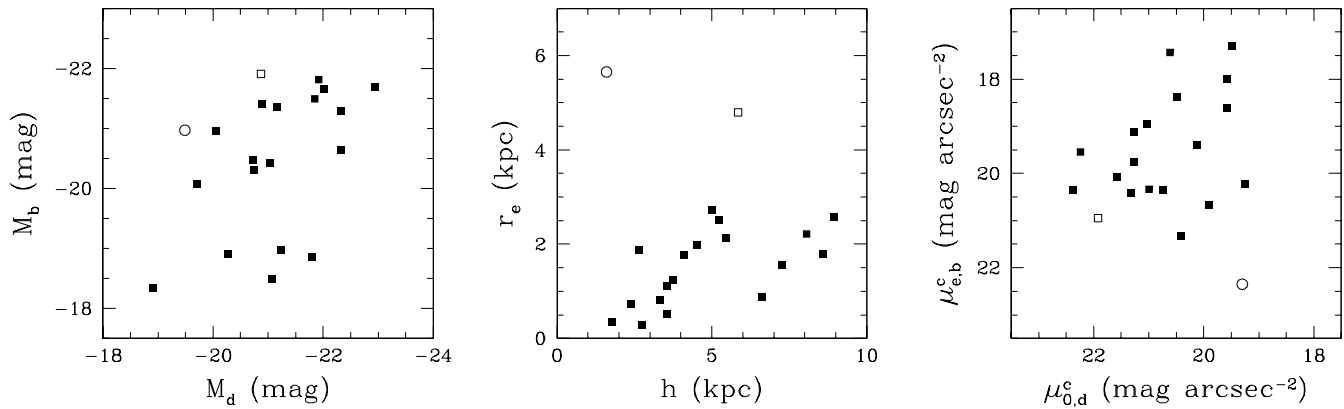,width=\textwidth}}
  \caption{Comparison of the R-band bulge and disk parameters. {\em
  Left:} absolute magnitudes of bulge vs.\ disk; {\em middle:} bulge
  effective radius vs.\ disk scale length; {\em right:} bulge
  effective surface brightness vs.\ disk central surface
  brightness. The open squares indicate UGC~624, the
  circles indicate UGC~6786.    
  \label{fig:bulgediskcomp}} 
\end{figure*}
The main purpose of the analysis described in this paper is to
derive the contribution of the stellar component of our galaxies to
their gravitational potential. 
In a forthcoming paper, we will calculate the contributions of the
stellar disks and bulges to the rotation curves, and use the results
to constrain the content and distribution of dark matter in our
galaxies.  
Meanwhile, some interesting results concerning the structure and mutual
dependence of the bulges and disks of our galaxies can be derived from
our study as well.
These results are in some ways complementary to the larger studies of 
e.g.\ \citet{Graham01}, \citet{MacArthur03}, \citet{DeJong04},
\citet{Hunt04} and \citet{Laurikainen05}. 
Below, we discuss the most important points.

\subsection{A comparison of bulge and disk parameters}
\label{subsec:bulgediskcomp}
In figure~\ref{fig:bulgediskcomp}, we compare the total luminosities, sizes
and surface brightnesses of the bulges and disks of the galaxies in our sample
with accurate bulge-disk decompositions. 
It is clear that there is only a weak coupling between the bulge and disk
parameters. 
 
There is a weak trend of more luminous disks harbouring more luminous
bulges, but the scatter around the relation is large. 
Moreover, several biases in our sample selection could introduce an
artificial correlation: galaxies with a highly luminous bulge and a
faint disk might not be classified as disk galaxies, but rather be
mis-identified as ellipticals.  
Similarly, luminous disks with faint bulges might not be classified as
early-type disks, but rather as late-type spirals. 
This latter effect is, however, not expected to be as strong as the
former, since the morphological type classification is based on other
parameters than bulge-disk luminosity ratios as well. 
Moreover, our sample contains a number of galaxies with
high-luminosity disks and faint bulges (e.g.\ UGC~94, 3205, 3546),
which are still clearly recognisable as early-type disk galaxies. 
UGC~12043 even has no obvious bulge component at all, but is still
classified as an S0/a galaxy. 

\begin{figure*}
  \centerline{\psfig{figure=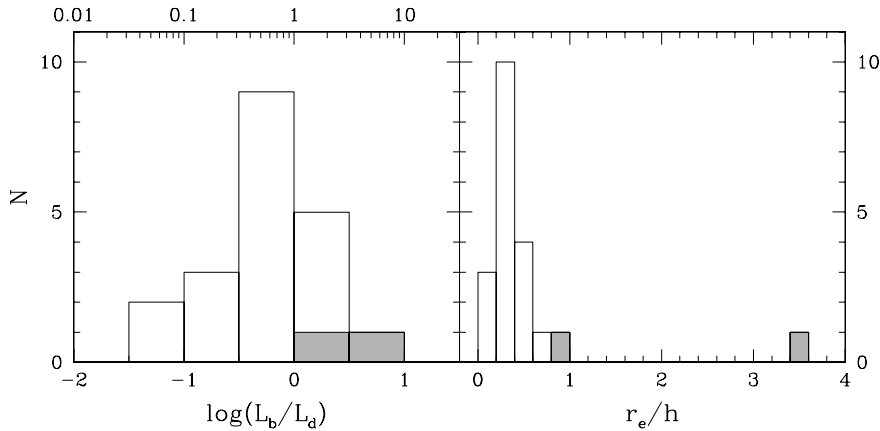,height=2.2in}} 
  \caption{The distribution of the R-band bulge-to-disk luminosity
  ratio ({\em left}) and size ratio ({\em right}) for the galaxies in
  our sample. Gray shading indicates UGC~624 and 6786. 
  \label{fig:BDhistos}}
\end{figure*}

In any case, our results indicate a large range in bulge-to-disk (BD)
luminosity ratio and show that the common belief that early-type disk
galaxies have large and luminous bulges does not hold in all cases.  
This is also visible in the left hand panel of
figure~\ref{fig:BDhistos}, where we show the distribution of 
the BD luminosity ratio for all galaxies in our sample. 
The average value of $\log(L_b/L_d)$ is $-0.23 \pm 0.47$ in the
R-band, where the error gives the standard deviation of the sample.   

Apart from two galaxies, UGC~624 and 6786, which have unusually large
bulges (as measured from their effective radii $r_e$), there is a weak
correlation between disk scale length and bulge effective radius
(middle panel in figure~\ref{fig:bulgediskcomp}).    
In the following, we will show that UGC~6786 differs from the rest of
the sample in many aspects, and we will interpret this in
section~\ref{sec:notes} as an indication that this galaxy is
not really a disk galaxy, but rather an elliptical galaxy with an 
additional disk of gas and stars.   
The offset of UGC~624 may be explained as a result of its high
inclination with respect to the line of sight. 
There are indications that the image of this galaxy is significantly
affected by dust (note for example the strong asymmetry in the centre,
visible in the $I_1/I_0$ lopsidedness parameter, and by eye in the
bulge-subtracted image). 
If there are indeed large amounts of dust present in this galaxy, it
is conceivable that they introduce large errors in the shape of the
photometric profiles and in the resulting bulge-disk decomposition.  

The other galaxies seem to follow the global trend that large bulges
reside in large disks, although the scatter is, again, large. 
In the right hand panel of figure~\ref{fig:BDhistos}, we show
the distribution of $r_e/h$ for all galaxies in the sample. 
The average value of this ratio is $0.51 \pm 0.72$ (the error
indicates the standard deviation), but this value is heavily
influenced by UGC~6786. 
The average value for the sample without UGC~6786 is $0.35 \pm 0.19$,
whereas it is further reduced to $0.32 \pm 0.15$ if UGC~624 is
excluded as well. 
Even this last value is significantly higher than the average $r_e/h$
for the late-type spiral galaxies studied by \citet{Courteau96}.
It is also higher than the values found by \citet{Graham01} and
\citet{MacArthur03}, who both noted a mild trend of the ratio $r_e/h$
increasing towards earlier type spiral galaxies, but only reaching an average 
of respectively 0.21 and 0.24 for the early types.  

It is not immediately clear what causes the higher $r_e/h$ ratio for
our sample, compared to the results of \citet{Graham01} and
\citet{MacArthur03}, but it could well be related to the different
decomposition techniques used.  
We already argued in section~\ref{subsec:bulges} that 1D 
decompositions suffer from systematic effects. 
Inspection of figure~\ref{fig:projection} on
page~\pageref{fig:projection} shows that, in an extreme case
such as UGC~6786, a photometric profile measured along ellipses with
ellipticity fixed at the value of the outer disk is shallower than the
true light distribution.
A simple least-$\chi^2$ 1D fit to this profile yields an effective
radius for this bulge of 24.0\arcsec, almost a factor 2 smaller than
the value we derived using our 2D method described in
section~\ref{subsec:bulges} (43.5\arcsec). 
Although UGC~6786 is an extreme galaxy, it seems likely that the same
effect plays a r\^ole, perhaps to a lesser extent, in other galaxies
as well.
Thus, the discrepancy between the average $r_e/h$ ratios of our study
and those of the 1D studies of \citet{Graham01} and
\citet{MacArthur03}, strengthens the need for 2D decompositions to
recover the bulge parameters accurately, especially in galaxies where
the bulges contribute a substantial fraction to the total light
\citep[cf.][]{Byun95, DeJong04}. 
Note that our average $r_e/h$ ratio is fully consistent with the study
of \citet{Khosroshahi00}, who used a 2D decomposition method for a
sample of predominantly early-type disk galaxies. 

No obvious trend is visible between the surface brightnesses of the
bulges and disks in our sample (right hand panel in
figure~\ref{fig:bulgediskcomp}, cf.\ \citealt{Hunt04}). 

\subsection{Correlations between different bulge parameters}
\label{subsec:bulgeprop}
\begin{figure*}
  \centerline{\psfig{figure=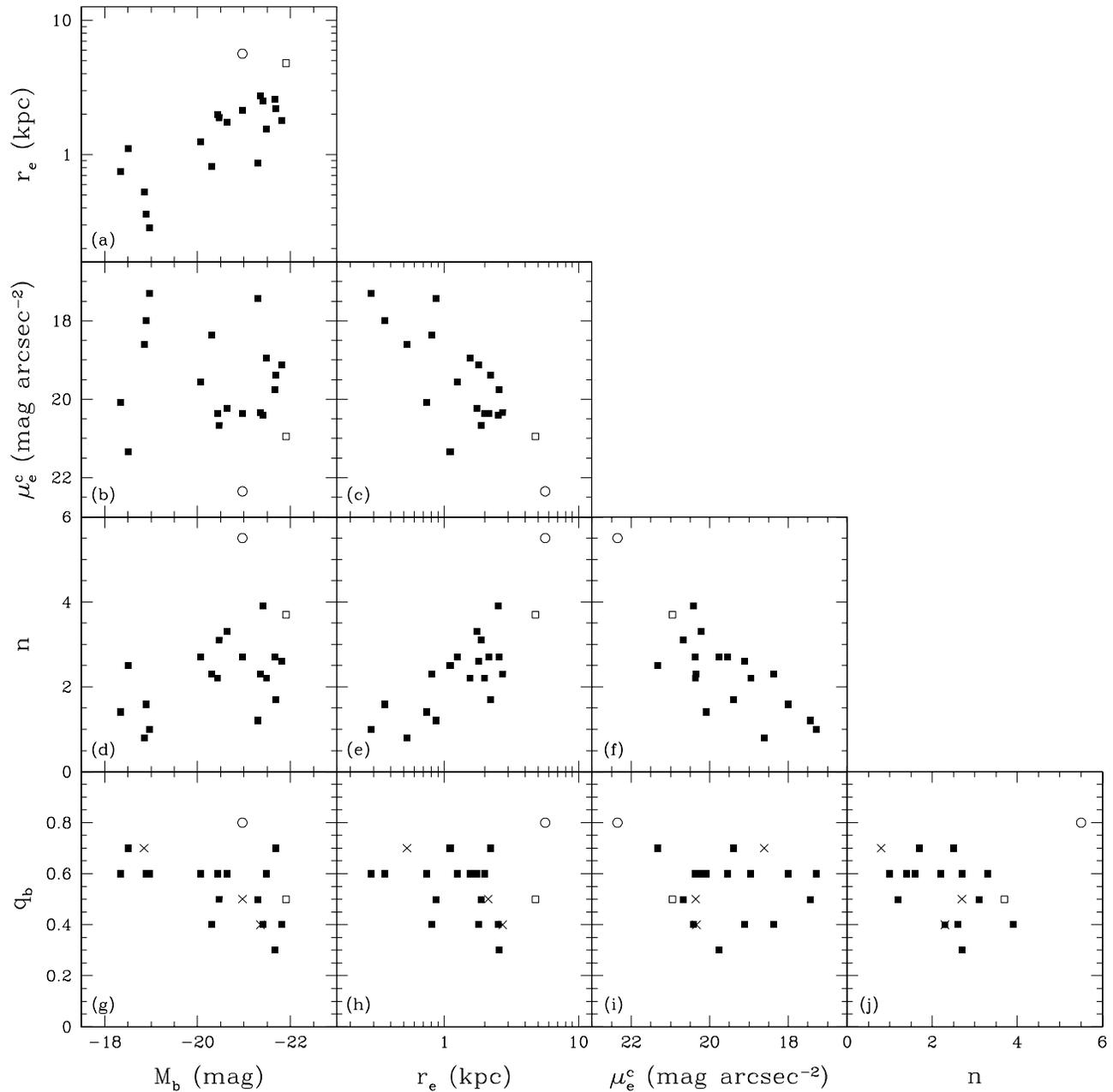,width=17cm}}
  \caption{Correlations between different bulge parameters: total
  magnitude $M_b$, effective radius $r_e$, effective surface
  brightness $\mu_e^c$ (corrected for Galactic foreground extinction),
  S\'ersic parameter $n$ and intrinsic axis ratio $q_b$. All data refer
  to the R-band images. The open squares indicate UGC~624, the circles
  indicate UGC~6786. The crosses in the bottom panels indicate bulges
  where determination of the intrinsic flattening of the bulge was
  difficult (UGC~2916, 3205 and 3993; see
  table~\ref{table:BDresults}).  
  \label{fig:bulgeprops}} 
\end{figure*}

Previous studies such as the ones by \citet{Khosroshahi00}, \citet{Graham01}
and \citet{DeJong04} have revealed several correlations between the different
parameters which characterise the bulges.  
In figure~\ref{fig:bulgeprops}, we study several of these
correlations for our galaxies. 

Panel (a) in this figure shows that the effective radius shows a clear
correlation with the total luminosity of the bulge. 
This is not very surprising, and simply shows that larger bulges are
more luminous. 

A priori less expected is the relation between the S\'ersic index $n$ 
and the total luminosity of the bulges (panel (d)): luminous bulges have
predominantly higher $n$-values than their low-luminosity
counterparts.  
This trend was noted before by e.g.\ \citet{Andredakis95},
\citet{Khosroshahi00} and \citet{Graham01}, and is also observed in 
elliptical galaxies \citep[e.g.][]{Caon93, Young94, Jerjen00}.  
It should, however, be approached with some caution, as there may be a
selection effect at play.  
Low-luminosity bulges are generally small (panel (a)) and in many
cases, their effective radii are only a few arcseconds on the sky (see
also table~\ref{table:BDresults}). 
In such cases, the seeing will smear out any sharp peaks in the light
profiles and, even though we deconvolve our bulge profiles when fitting
the S\'ersic model, the $n$-value will effectively be lowered
\citep{Trujillo01b}. 
Thus, the apparent correlation in panel (d) may be partly artificial,
and sub-arcsecond images would be required to confirm the trend at the
low-luminosity end \citep[cf.][]{Balcells03}.  

\begin{figure}
 \centerline{\psfig{figure=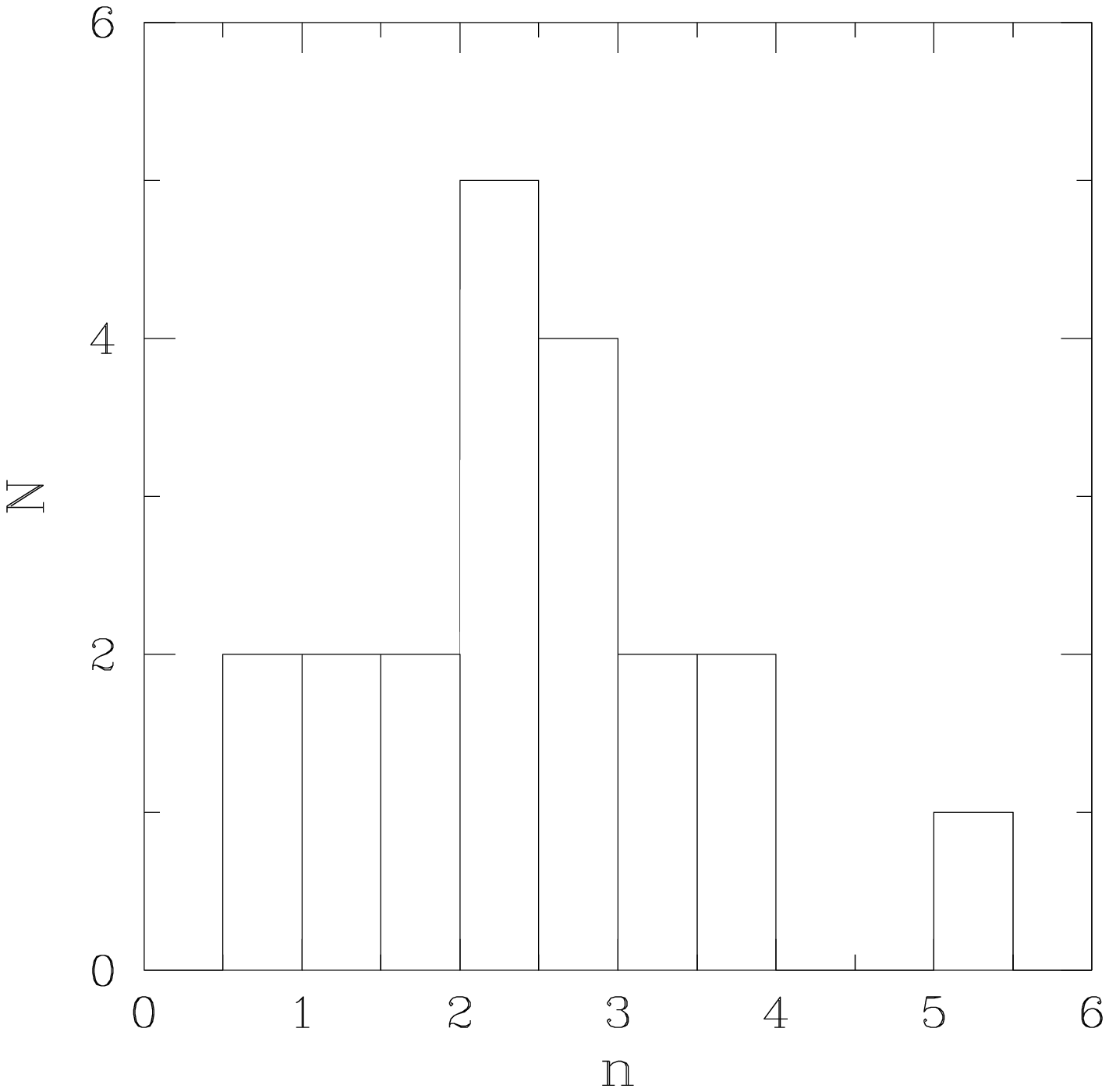,width=7.5cm}}
 \caption{The distribution of the S\'ersic $n$-parameter for the
  bulges in our sample. The numbers were derived from the R-band 
  data. \label{fig:nhisto}}
\end{figure}

The bulges in our sample span the full range from exponential light
profiles ($n \approx 1$) to De Vaucouleurs ($n \approx 4$), with
UGC~6786 lying out at $n = 5.5$. 
A histogram of the distribution is shown in
figure~\ref{fig:nhisto}. 
The average value of $n$ for the entire sample is $2.5 \pm 1.1$,
where the error gives the standard deviation of the sample; for the
sample without UGC~6786, this becomes $\langle n \rangle = 2.3 \pm
0.9$.  
These values are fully consistent with previous results from e.g.\
\citet{Andredakis95, DeJong96, Graham01} and \citet{Hunt04} and
confirm the view that the bulges of early-type disk galaxies form, at 
least as far as their $n$-values is concerned, an intermediate
population between elliptical galaxies, which have $n$ around 4 but up
to 15 in extreme cases \citep{Caon93, DeJong04}, and late-type galaxy
bulges, which usually have exponential or even steeper profiles
\citep[e.g.][]{DeJong96, MacArthur03}.    

The three fundamental bulge parameters from equation~\ref{eq:Sersicmag},
$\mu_e$, $r_e$ and $n$ are all clearly correlated (panels (c), (e) and (f) in 
figure~\ref{fig:bulgeprops}, cf.\ \citealt{Khosroshahi00}, \citealt{DeJong04},
\citealt{Hunt04}), but the selection effect described above may be important
here as well: seeing effects may introduce a bias for small bulges towards
lower $n$. 
Furthermore, the correlations may partly be caused by the definition 
of these parameters which intrinsically couples them.  
Bulges with higher $n$ values have a shallower light profile at large
radii, but with a suitable choice of the central surface brightness
and effective radius, the profile can look relatively similar to that
of low-$n$ bulges in the inner parts. 
Such high-$n$ bulges will have a larger effective radius $r_e$ than
low-$n$ bulges, and the surface brightness $\mu_e$ at that radius will
be lower. 
These are exactly the trends which are visible in
figure~\ref{fig:bulgeprops}. 

\begin{figure}
 \centerline{\psfig{figure=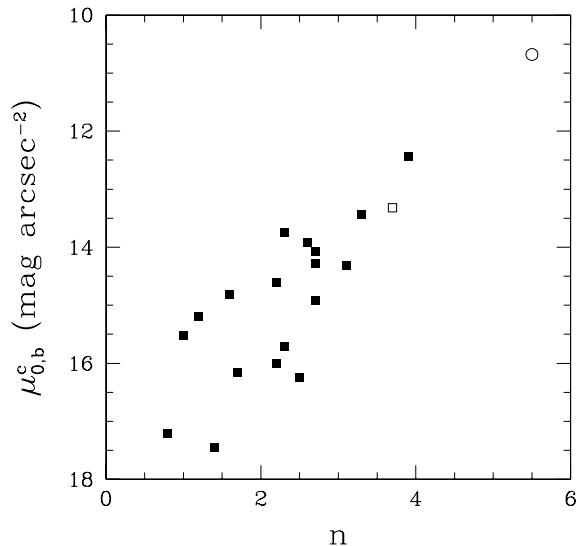,width=7.5cm}}
 \caption{The correlation between S\'ersic $n$-parameter and fitted
  central surface brightness $\mu_0$ (corrected for Galactic
  foreground extinction) for the bulges in our sample. All data are
  derived from the R-band images.\label{fig:mu0plot}}  
\end{figure}
Parameter coupling alone cannot, however, explain the correlations 
between the bulge parameters. 
This can be appreciated, for example, by considering instead of the
effective surface brightness the {\em central} surface brightness
$\mu_0$. 
In figure~\ref{fig:mu0plot}, we show that the fitted central
bulge surface brightness is highly variable, with a similar width in
the distribution as for the effective surface brightness $\mu_e$.
Moreover, it is strongly correlated with the S\'ersic shape parameter
as well.  
Thus, for a given $n$, bulges occupy a distinct region in
surface brightness space.
The correlations between the bulge parameters are therefore real and
must have a true physical basis \citep[cf.][]{Trujillo01}.  

There seems to be a weak trend between the flattening of the bulges
and their total luminosity, more luminous bulges being on average
more flattened than low-luminosity bulges (panel (g) in
figure~\ref{fig:bulgeprops}).  
This result is somewhat surprising, since previous studies had found 
that flat `pseudobulges' are more common in late-type spiral galaxies
than in early-type disk galaxies \citep[][ and references
therein]{Kormendy93, Kormendy04}. 
One would thus expect the most spherical bulges in our sample to be
the most luminous, but this is clearly not the case.

It is important to note that the uncertainties in individual
measurements of the intrinsic axis ratio of the bulge are large.  
In several cases, the ellipticity of the bulge isophotes is not
constant with radius, and it is problematic to assign a unique value
to the apparent axis-ratio of the bulge. 
In other cases, where the bulges are small, seeing effects may lead to
biases as well. 
Finally, in galaxies that are close to face-on, there is little
leverage on the problem, as different intrinsic axis ratios lead to
very minor changes in the observed image.   
The three galaxies where the determination of the bulge flattening was
particularly difficult (UGC~2916, 3205 and 3993) are indicated with
crosses in the bottom panels in figure~\ref{fig:bulgeprops}. 
In spite of all these problems, the correlation in panel (g) is
suggestive, and the uncertain data points do not appear to alter it.
There seems to be a genuine trend of more luminous bulges being
flatter than less luminous ones. 

\begin{figure}
 \centerline{\psfig{figure=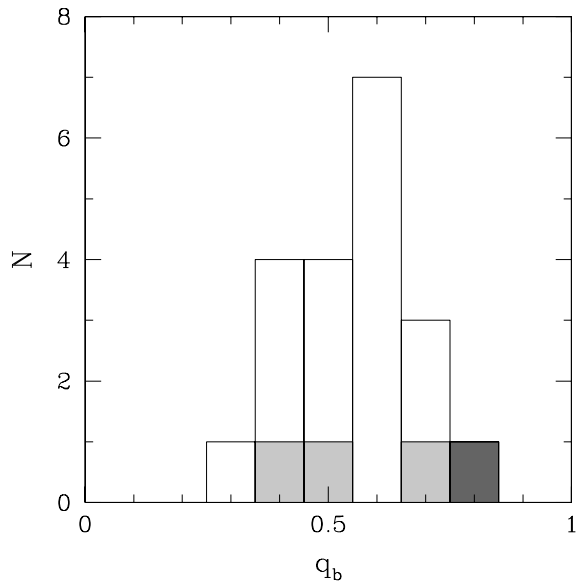,width=7.5cm}}
 \caption{The distribution of the intrinsic axis ratios $q_b$ of the 
  bulges in our sample. All values were derived from the R-band
  data. The white histogram shows the distribution for the entire
  sample, the light gray shading indicates galaxies where the
  determination of $q_b$ was particularly difficult (UGC~2916, 3205 and
  3993, see table~\ref{table:BDresults}). Dark shading indicates
  UGC~6786.\label{fig:qhisto}}    
\end{figure}
Finally, the distribution of the intrinsic axis ratios $q_b$ for our 
entire sample of bulges is shown in figure~\ref{fig:qhisto}.  
The average value for the sample is $\langle q_b \rangle = 0.55 \pm
0.12$, where the error gives the standard deviation of the sample.   
Exclusion of the three uncertain values does not lead to different
values. 
Note that none of the galaxies in our sample harbours a truly
spherical bulge. 
Furthermore, the least flattened bulge is that of UGC~6786, which
differs from the rest of the sample in many respects (see discussion
above) and seems to be more resemblant of an elliptical galaxy with a
small disk of stars and gas (see also the note in
section~\ref{sec:notes}). 
All other bulges have an intrinsic axis ratio of 0.7 or less. 

\subsection{Colour gradients}
\label{subsec:colourgradients}
\begin{figure*}
  \centerline{\psfig{figure=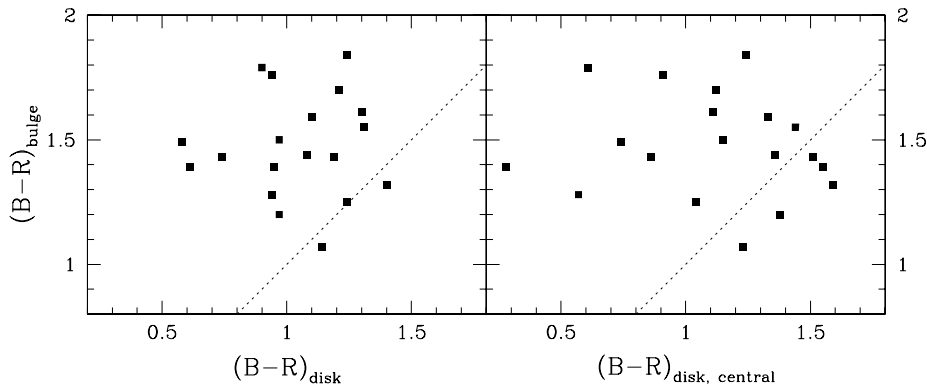,height=5.cm}}
  \caption{Bulge B-R colour vs.\ disk colour. In the left panel, the
  bulge colour is plotted vs.\ the integrated disk colour; in the
  right panel, it is plotted against the fitted central disk colour, 
  $\mu_{0,B}^c - \mu_{0,R}^c$. The dashed line gives the locations of
  equal colour. All colours are corrected for Galactic foreground
  extinction. \label{fig:bulgediskcols}} 
\end{figure*}
It has long been known that many spiral galaxies do not have a uniform
colour over their entire disk, but instead become bluer towards larger
radii \citep{DeVaucouleurs61, DeJong96b, Matthews99}. 
Such colour gradients have been interpreted in the past as a result of
radial variations in the dust content and star formation history (SFH)
in the disks of spiral galaxies.
\citet{DeJong96b} concluded, based on detailed radiative transfer and stellar
synthesis models, that dust reddening plays only a minor r\^ole and that most
of the colour gradients are due to ``a combined stellar age and metallicity
gradient across the disk, with the outer regions being younger and of lower
metallicity''. 
On the other hand, \citet{Mollenhoff06} recently argued that ``the tendency
for apparent scalelength to increase with decreasing wavelength is primarily
due to the effects of dust''. 
Clearly, the underlying mechanisms causing the colour gradients are not fully
understood yet.

Here we investigate whether colour differences between the bulges and
the disks are responsible for (part of) the observed colour gradients
in our galaxies.

Inspection of the figures in appendix~\ref{app:atlas} shows
that many galaxies in our sample have colour gradients; most galaxies
become bluer towards larger radii.
There are, however, also galaxies which show no evidence for colour
variations, or which become redder towards large radii (e.g.\ UGC~89,
2953).  
It is important to note that the errorbars on the colour profiles
become large in the outer regions of the galaxies, predominantly as a
result of the uncertainties in the determination of the background
level of the images. 
Although we attempted to determine the background level in our images as
accurately as possible and we tried to include its uncertainty in the
errorbars of our photometric profiles (see section~\ref{subsec:surfphot}), it
is possible that the large colour variations in the outer parts of e.g.\
UGC~6787 are the result of incorrect background levels in one of the colour
bands.   
Colour variations in the bright inner regions of the galaxies cannot
be explained by an improper background determination and must be
real. 

The results from our bulge-disk decompositions indicate that the
bulges of our galaxies are on average significantly redder than the
disks surrounding them (see figure~\ref{fig:bulgediskcols}).  
The average B-R colour difference between the bulges and disks is
$\langle (B-R)_{\mathrm {bulge}} - (B-R)_{\mathrm {disk}} \rangle =
0.43 \pm 0.30$, where the error gives the standard deviation of the
sample.  
This colour difference between bulges and disks is, in most cases, at least
partly responsible for the observed colour gradients: the relative
contribution of the red bulge light declines towards larger radii, at the cost
of an increasing contribution of bluer disk light. 
Good examples of this effect are UGC~94, 1541, 5253 and 11670.

However, the colour differences between bulges and disks are not sufficient to
explain all radial colour gradients. 
In some cases, the bulge-subtracted images (shown at the bottom right
in the figures in appendix~\ref{app:atlas}) do not contain
any remaining colour variations, showing that indeed the colour
gradients in the original images were caused by the influence of the
bulges (e.g.\ UGC~2916, 9133, 11670). 
But in most cases, the bulge-subtracted images still contain colour
gradients and become bluer towards larger radii (e.g.\ UGC~2487, 3205,
3993, 6787). 
Thus, our results show that not only are bulges on average redder than
their surrounding disks, the disks themselves are also bluer in the
outer parts than at smaller radii. 
If De Jong's conclusion was correct that dust reddening plays only a
minor r\^ole in the colour gradients \citep{DeJong96b}, then our
results indicate that the stars in the bulges were on average formed
earlier than those in the disk (or have higher metallicity), and that
in the disks themselves, the star formation history and metallicity
vary with radius (with the stars in the outer regions being younger
and/or having lower metallicity).

\section{Notes on individual galaxies}
\label{sec:notes}
{\setlength{\parskip}{0cm}
{\bf UGC~89} (NGC~23) has a relatively small ($r_e/h = 0.13$) but bright bulge
(central surface brightness $\mu_{c,R} = 15.9 \magasas$). The presence of the
strong bar complicates the determination of the bulge parameters, as it
dominates the light distribution to very small radii. In particular, the
ellipticity of the isophotes does not reach a constant value in the bulge
region, so we were forced to estimate the bulge ellipticity $\epsilon_b$ by
eye. The bar is also responsible for the characteristic `dip' around $r =
35\arcsec$ in the surface brightness profiles. \\[1.5mm]   
{\bf UGC~94} (NGC~26) has a very low bulge-to-disk luminosity ratio of
$L_b/L_d = 0.09$ in R (0.04 in B). The low luminosity of the bulge is
caused by its low effective surface brightness ($\mu_e^c = 21.4\magasas$ in
the R-band), while its effective radius is normal ($r_e/h = 0.31$). Of all
bulges in our sample, this one lies furthest to the bottom-left in panel (b)
of figure~\ref{fig:bulgeprops} ($\mu_e^c$ vs.\ $M_b$). Although UGC~94 is not
classified as a barred galaxy, it has a  small but distinct bar in the centre,
oriented approximately east--west. \\[1.5mm]     
{\bf UGC~624} (NGC~338) has a relatively large and luminous bulge
($r_e/h = 0.82, L_{b,R}/L_{d,R} = 2.60$, see
figure~\ref{fig:bulgediskcomp}). These extreme properties
may, however, be a result of the poor bulge-disk decomposition for
this galaxy. The combination of an exponential disk and S\'ersic bulge 
seems a rather inadequate description of this galaxy and the extreme
bulge properties may be simply an artefact caused by the fitting of an
inappropriate model to the data. Part of the problem with UGC~624 may
be the high inclination and resulting dust extinction. There are
strong indications for dust extinction in the centre (note for example
the strong asymmetry in the centre, visible in the $I_1/I_0$
lopsidedness parameter, and by eye in the bulge-subtracted image). It
is not inconceivable that the peculiar shape of the luminosity profile
of this galaxy is at least partly caused by dust extinction. \\[1.5mm]  
{\bf UGC~1541} (NGC~797) has a distinct bar which causes the typical
`bump' in the luminosity profile around 40\arcsec. To the west, at the
end of the spiral arm, a small companion elliptical galaxy is visible 
\citep[cf.][]{Zwicky71}. The light of this companion was masked out
for the measurement of the luminosity profile and the subsequent
bulge-disk decomposition.
 
The R-band image of this galaxy, which is shown in
the figure in the appendix, was taken in windy conditions. As a result, the
telescope suffered from mild vibrations, causing the PSF to be elongated,
roughly along the east-west direction. To avoid errors due to this effect, we
measured the position angle and ellipticity of the bulge from the B-band
image. Outside the region affected by the PSF elongation, the R-band image is
consistent with the bulge orientation thus derived. \\[1.5mm]   
{\bf UGC~2487} (NGC~1167) is the most luminous galaxy in our sample 
($M_{\mathrm{lim},R} = -23.24$). It is an almost perfect superposition  
of an exponential disk and a S\'ersic bulge; the residuals with 
respect to the fitted bulge and disk profiles are less than
0.05\magasas\ everywhere. Some very faint spiral structure is visible
in the disk. \\[1.5mm]    
{\bf UGC~2916} has a nuclear bar; the ellipticity of the isophotes in
the centre does not represent the shape of the bulge and we were
forced to estimate the bulge ellipticity $\epsilon_b$ by eye. As as
result, the derived intrinsic axis ratio of the bulge is
uncertain. The colour gradient in this galaxy can be fully explained
by the colour difference between the bulge and disk; the colour of the 
bulge-subtracted image does not vary with radius. \\[1.5mm] 
{\bf UGC~2953} (IC~356) is a large and well-resolved galaxy on the
sky, but it is relatively nearby ($D = 15.1$~Mpc) and the physical
dimensions are rather average for our sample ($h =$~4.1~kpc). The
decomposition into exponential disk and S\'ersic bulge is almost
perfect, with residuals less than 0.1\magasas\ almost everywhere.  It
is also one of the most symmetric galaxies in our sample, without any
strong lopsided or oval distortions. \\[1.5mm] 
{\bf UGC~3205} has the lowest R-band bulge-to-disk luminosity ratio of
all galaxies in our sample ($L_b/L_d = 0.07$). Its bulge is also very
small compared to the surrounding disk ($r_e/h = 0.15$) and is only
barely resolved in our images ($r_e = 2.2\arcsec$, 1.7 times the
seeing of the R-band observations). As a result, the fitted S\'ersic
parameter $n$ and intrinsic axis ratio $q_b$ for this bulge have a
limited accuracy; higher-resolution observations are necessary to
study the structure of the bulge of this galaxy in detail. Although
not classified as a barred galaxy, UGC~3205 has a weak bar, roughly
oriented east-west (see also the $I_2/I_0$ Fourier term in the figure
in the appendix). Apart from the bar, the galaxy is highly
symmetric. No spiral structure can be seen in our images; a
classification as S0 seems more appropriate than the Sab from the
UGC. \\[1.5mm]   
{\bf UGC~3546} (NGC~2273) has the smallest bulge of all galaxies in
our sample, both in absolute terms ($r_e = 0.29$~kpc) as relative to
the surrounding disk ($r_e/h = 0.10$), but due to its high surface 
brightness, the bulge-to-disk luminosity ratio is larger than in
UGC~3205 ($L_b/L_d = 0.12$). The bulge is embedded in a distinct bar;
two spiral arms emanate from the ends of it and form a ring in the 
outer parts \citep[cf.][]{VanDriel91b}. In the very inner parts, a
secondary bar seems present, but high-resolution observations with HST
by \citet{Erwin03b} showed that it is actually a nuclear ring with a
star-forming spiral inside. UGC~3546 has a Seyfert~2 nucleus
\citep{Huchra82} and has been extensively studied at radio wavelengths
\citep[e.g.][]{Ulvestad84, Nagar99}. This galaxy was observed during
three different runs in R and two runs in B, enabling a thorough
check on the internal consistency of our observations and data
reduction techniques (see section~\ref{sec:comparison}); all 
observations were found to be fully consistent. \\[1.5mm]
{\bf UGC~3580} is a relatively small and low-luminosity galaxy, with
an absolute R-band magnitude of -19.42 and an R-band disk scale length
of 2.4~kpc. It has an irregular appearance and seems strongly
affected by dust. Although there is clearly excess light in the centre
over the exponential outer disk, this `bulge' has a patchy light
distribution, bearing little resemblance to the smooth spheroids
present in most other galaxies in our sample. It is also bluer than
the other bulges in our sample ($(B-R)_b = 1.05$, corrected for
Galactic foreground extinction). Several `knots' of bright emission
are seen in the image as well, presumably regions of active star
formation; \citet{James04} report an equivalent width for \Ha + \NII\
of 2.9~nm, unusually large for an early-type spiral galaxy, indicating
that indeed, this galaxy is actively forming stars. In a forthcoming
paper, we will show that the kinematical structure of this galaxy is
also different than that of the other galaxies in our sample.

All these facts lead us to assume that UGC~3580 is a (relatively
luminous) member of a class of dwarf early-type spiral galaxies, which
have distinctly different morphological and kinematical properties
than the more common high-luminosity early-type disk galaxies. Other 
galaxies in our sample which probably belong this class are UGC~12043 (see 
below), and UGC~6742 and 12713 \citepalias{Noordermeer05}. \\[1.5mm]
{\bf UGC~3993} is an example of a luminous bulge embedded in a
low surface brightness (LSB) disk. The fitted central surface
brightness of the disk is 23.2\magasas\ (B-band, corrected for
Galactic foreground extinction and inclination effects), but the
description as an exponential disk is rather poor and this value may
not represent the actual surface brightness of the disk in the
centre. The bulge-subtracted image in the appendix shows that the disk
is dominated by an inner ring with a radius of approximately
10\arcsec\ ($\sim 3$~kpc). The exact density contrast of this ring with the
surrounding disk regions depends quite sensitively on the details of the
bulge-disk decomposition; in particular, acceptable decompositions can be
performed where the central hole in the disk is not as deep as in our
preferred solution. In all solutions, however, there is a clear overdensity at
this radius, showing that the ring is a genuine physical feature. In addition
to this inner ring, some very faint, filamentary spiral arms are visible in
the outer regions of the disk.  

Due to the near face-on orientation of this galaxy ($i \approx 23\deg$), the
bulge and disk ellipticity could not be distinguished, and the bulge
flattening could only be guessed. \\[1.5mm]   
{\bf UGC~4458} (NGC~2599) has the second most luminous bulge of all
galaxies in our sample ($M_{b,R} = -21.8$), surpassed only by that of
UGC~624 (but see note for that galaxy above). The surrounding disk is
highly extended, with an R-band exponential scale length of 8.6~kpc
(second largest of all galaxies in our sample, after UGC~9133). Some
filamentary structures can be recognised on the northwest of the disk,
either faint spiral arms or shells caused by a merging event.\\[1.5mm]
{\bf UGC~5253} (NGC~2985) has a large featureless bulge which is well
fitted with an $R^{1/4}$-profile (we find $n = 3.9$). However, the
ellipticity of the isophotes in the bulge-dominated regions is hardly
lower than in the outer, disk-dominated regions, indicating that the
bulge is highly flattened; we derive an intrinsic axis-ratio of
$q_b=0.4$. The disk is dominated by beautiful, grand design spiral arms;
many `knots' of bright blue emission can be discerned, presumably star
forming regions \citep[cf.][]{Delgado97}.\\[1.5mm]  
{\bf UGC~6786} (NGC~3900) is in many ways different from the other galaxies in
our sample (cf.\ section~\ref{sec:discussion}). It has the largest and most
luminous bulge of all galaxies, compared to the disk ($L_b/L_d = 3.91$, $r_e/h
= 3.54$, all in R) and the S\'ersic parameter and intrinsic axis ratio of the
bulge are larger than in any other bulge in our sample ($n = 5.5$, $q_b =
0.8$). The bulge is so dominant that the isophotes of the original image do
not probe the disk elongation at any radius, so the inclination derived using
the standard method (section~\ref{sec:obs_and_redux}) was not correct. Only
after the subtraction of the bulge could the disk be distinguished properly
and could we derive its orientation from the shape of the faint spiral
structure. For the derivation of the disk profile and the intrinsic axis ratio
of the bulge, we assumed an ellipticity of 0.21, corresponding to an
inclination of 69\deg\ (close to the kinematical inclination angle of 68\deg,
derived from the \HI\ velocity field).   

The resulting disk profile shows a plateau of constant surface density
in the centre ($\mu_R \approx 21.5\magasas$) and an
exponentially decaying profile outside 35\arcsec\ ($\sim
4.5$~kpc). A linear fit to the outer parts of the disk profile yields
a scale length of 12.3\arcsec\ (1.6~kpc) and a formal central surface
brightness of 19.3\magasas\ (R-band, corrected for Galactic
foreground extinction and inclination). Note, however, that such high surface
brightness is in reality not reached in the disk, due to the truncation of the
profile around 35\arcsec. 

All combined, UGC~6786 does not look like a normal early-type disk galaxy at
all, but rather seems to resemble an elliptical galaxy which has acquired a
disk of gas and stars. In the recent past, it was discovered that many
galaxies which are classified as ellipticals, harbour small stellar disks
\citep[e.g.][]{Rix90, Scorza95}; UGC~6786 could well be an extreme example of
such a class of galaxies. There may also be a connection with galaxies such as
NGC~3108 or NGC~4278, elliptical galaxies with extended, regularly rotating
gas disk around them \citep{Oosterloo02, Morganti06}. It is conceivable that
the gas in such galaxies will eventually form stars as well, and in fact,
there is some evidence for a very faint stellar disk in the centre of
NGC~3108. UGC~6786 may be a similar, but more evolved, system to these
galaxies.\\[1.5mm]      
{\bf UGC~6787} (NGC~3898) suffers from the same problem as UGC~6786:
the luminous bulge distorts the isophotes of the disk out to large
radii and the standard method to derive the inclination
(section~\ref{subsec:isoorient}) cannot be used. Only after
the model image of the bulge was subtracted could the disk ellipticity
be determined. The final disk photometric profile and the intrinsic
axis ratio of the bulge were derived using an inclination of
61\deg. Note that this is still somewhat lower than the kinematic
inclination which we derived from the \HI\ velocity field
(69\deg). \\[1.5mm]      
{\bf UGC~8699} (NGC~5289) is highly inclined ($i = 77\deg$), but we
can still recognise the bright bulge, small bar and surrounding
ring. The disk component in this galaxy has a low surface brightness
($\mu_{0,d}^c = 23.4 \magasas$, B-band, corrected for inclination
effects and Galactic foreground extinction). The strong asymmetry in
the central light distribution is most likely the result of internal
absorption by dust; at larger radii, the galaxy is highly
symmetric. The sudden change in position angle and ellipticity around
70\arcsec\ is a result from imperfect flatfielding and is not
real. \\[1.5mm]     
{\bf UGC~9133} (NGC~5533) is a luminous galaxy ($M_R = -22.62$) which
can almost perfectly be decomposed in an exponential disk and S\'ersic
bulge (the residuals being less than 0.1~mag arcsec$^{-2}$ almost everywhere).
It has the largest disk scale length of all galaxies in our sample
($h_R = 8.9$~kpc). The ellipticity of the bulge is barely lower than
that of the disk, implying that the bulge is highly flattened; we
derive an intrinsic axis-ratio of $q_b = 0.3$. Some faint, filamentary
spiral structure is visible in the outer disk. \\[1.5mm]  
{\bf UGC~11670} (NGC~7013) is a small and relatively low-luminosity
galaxy ($h_R = 1.8$~kpc, $M_B = -19.20$). It has a large, lens-shaped
bar with conspicuous dust-lanes running parallel to it. In the outer
parts, a diffuse disk with very faint spiral structure can be
seen. \\[1.5mm]   
{\bf UGC~11852} is the most distant galaxy in our sample ($D =
80$~Mpc). It consists of a relatively diffuse bulge embedded in a
highly elongated bar. The spiral arms are somewhat irregular and
develop into narrow filaments extending to very large radii (barely
visible in the figure in appendix~\ref{app:atlas}). \\[1.5mm]  
{\bf UGC~11914} (NGC~7217) is a highly symmetric, nearly face-on
galaxy, which can be well decomposed into a S\'ersic bulge and an
exponential disk. The bulge is relatively large, compared to the disk
($r_e/h = 0.71$). The residuals with respect to the model components
are small, except for a distinct ring of blue stars which causes
excess light at a radius of approximately 75\arcsec\ \citep[$\approx 
5.5$~kpc; cf.][]{Buta95, Verdes-Montenegro95}. This ring 
coincides with an enhanced surface density of neutral gas
\citepalias{Noordermeer05} and contains many star-forming regions
\citep{Pogge89b, Battinelli00}. However, as discussed in  
\citetalias{Noordermeer05}, the gas surface densities are below
the star-formation threshold from \citet{Kennicutt89}, so it is
puzzling how this galaxy can sustain its large-scale star formation
activity. 

Note that our bulge-to-disk luminosity ratio for UGC~11914 is much
lower than the value derived by \citet{Buta95} (0.58 vs.\ 2.3, both 
B-band); this difference is presumably caused by the fact that they
used an $R^{1/4}$-profile for the bulge, whereas we left the S\'ersic
index free in the fits. Our fitted bulge profile has $n = 3.1$, and
thus contains much less light at large radii than an
$R^{1/4}$-bulge. This difference illustrates once more the importance of
using a general S\'ersic profile for the bulge intensity distribution,
rather than the less flexible de Vaucouleurs profile. \\[1.5mm]    
Although {\bf UGC~12043} (NGC~7286) has the characteristic smooth
light distribution of early-type disk galaxies, it completely lacks a
central bulge component. The light profile shows only an exponential
disk; the fitted exponential profiles are overplotted with the bold
lines in the figure on page~36. This
galaxy is also smaller and less luminous than other galaxies in
our sample ($D_{25,R}^c = 7.8$~kpc, $M_R = -18.26$). This galaxy seems
a member of a class of dwarf early-type spiral galaxies with
distinctly different properties from their high-luminosity early-type 
counterparts (see also UGC~3580). \\[1.5mm]  
}

\section{Summary and Conclusions}
\label{sec:conclusions}
We have obtained deep R- and B-band surface photometry for a sample of 
21 early-type disk galaxies with morphological types between S0 and
Sab and B-band absolute magnitudes between -17 and -22. 
For 6 galaxies, I-band photometry is available as well. 
On average, our data reach surface brightness levels of 26.95, 26.07 and
24.26\magasas\ ($3 \sigma$) in B, R and I respectively.

For all galaxies, we have presented the results of our isophotal
analysis, including radial variations of surface brightness, colour, 
ellipticity, position angle and deviations from axisymmetry.   
We have also determined isophotal and effective diameters and total
magnitudes. 

We have developed a new, interactive, bulge-disk decomposition method
which takes into account the full 2D information from the images and
decomposes them into spheroidal bulges with a general S\'ersic intensity
profile and disks with an arbitrary intensity distribution. 
We made no prior assumptions about the intrinsic axis ratios of the bulges,
but rather determined those directly from the images by comparing the
ellipticity of the bulge isophotes to that of the outer disk.   
We have shown that 1D bulge-disk decomposition methods, where analytic
functions are fit directly to the radial photometric profiles, suffer from 
systematic biases, particularly in galaxies with large and luminous
bulges. 
For example, they can lead to severe under-estimates of the effective radii of
the bulges. 
2D techniques, which use the full information available in the image, are less
affected by such biases and yield more accurate results on the structural
parameters of the bulges. 

From a comparison of different bulge and disk parameters, we find the
following results concerning the structure of early-type disk galaxies:
\begin{list}{--}{\leftmargin=0.5cm \itemsep=0.1cm \parsep=0cm
\topsep=0.1cm} 
 \item There is a wide range in bulge-to-disk luminosity and size
 ratios. The average value of $\log(L_b/L_d)$ is $-0.23 \pm 0.47$ and 
 the average ratio $r_e/h$ is $0.32 \pm 0.15$ (excluding two galaxies,
 UGC~624 and 6786, which have unusually large bulges, see
 section~\ref{sec:notes}). The errors give the standard
 deviations of the samples. The common belief that early-type disk
 galaxies have large and luminous bulges does not hold in all cases;
 our sample contains several galaxies with faint and/or small bulges
 compared to the surrounding disks (e.g. UGC~94, 3205, 3546). 
 \item Luminous bulges have on average larger values for the S\'ersic
 shape parameter $n$, consistent with previous studies. The average
 value for all bulges in our sample is $\langle n \rangle = 2.5 \pm
 1.1$. Bulges of early-type disk galaxies form, at least as far as
 their $n$-values is concerned, an intermediate population between
 elliptical galaxies and late-type galaxy bulges.  
 \item The three fundamental bulge parameters, $\mu_e$, $r_e$ and $n$
 are all correlated, with the large bulges having a lower effective
 surface brightness and a more strongly centrally peaked light
 distribution (higher $n$). 
 \item None of the galaxies in our sample harbours a truly spherical bulge; in
 contrast, several bulges are highly flattened with intrinsic axis ratios as
 low as 0.3 -- 0.4 (e.g.\ UGC~5253, 9133). The average intrinsic axis ratio of
 the bulges in our sample is $\langle q_b \rangle = 0.55 \pm 0.12$. The
 flattening of the bulges is weakly coupled to their luminosity, more luminous
 bulges generally being more flattened than low-luminosity systems. The
 scatter in the relation is, however, large and more observations are needed
 to investigate this trend further. 
 \item The fact that most bulges in our sample are significantly flattened and
 have an intensity profile shallower than $R^{1/4}$ suggests that many of the
 bulges in our sample are not miniature elliptical galaxies, but rather 
 `pseudobulges' formed from disk material by secular processes, in accordance
 with the ideas proposed by \citet{Kormendy04}. However, whereas they assumed
 that such pseudobulges occur mainly in later-type spiral galaxies, our
 results imply that they may be common in massive, early-type disk galaxies as
 well \citep[cf.][]{Laurikainen06}. 
 \item The single exception to the last point is UGC~6786. This galaxy
 is fully dominated by its bulge ($L_b/L_d = 3.91$, $r_e/h = 3.54$,
 all in R), which has a S\'ersic $n$-parameter of 5.5 and is almost
 spherical ($q_b = 0.8$). It has a clear disk component, but its
 luminosity profile is unusual, with a plateau of constant
 surface-density in the centre and a steep exponential fall-off in the 
 outer parts. It seems most logical to interpret this galaxy as an
 elliptical which has later acquired a disk of gas and stars (e.g.\
 due to accretion or a merger event).  
 \item Many galaxies become bluer towards larger radii. In some
 cases, this can be explained solely by the radially declining
 contribution of the red bulge to the total light. In most cases,
 however, this effect is not sufficient and the disks themselves must
 contain colour gradients as well. 
\end{list}

The results presented in this paper will be used in a future paper to
calculate the contributions of the stellar components to the rotation
curves of these early-type disk galaxies.

\section*{Acknowledgments}
EN is grateful to Alister W. Graham for helpful discussions about
various issues related to bulge-disk decompositions, and for kindly
providing the FORTRAN code to perform the least-squares fits to the
bulge profiles.  
We would also like to thank Jelte de Jong for carrying out the
observations of UGC~6786 and 7989 on the 2.4m MDM Hiltner Telescope. 
We thank the referee, Phil James, for pointing out a few unclarities in the
original version of the manuscript and for several suggestions for
improvement.

\bibliographystyle{mn2e}
\bibliography{../../../references/abbrev,../../../references/refs}

\begin{thebibliography}{}

\bibitem[\protect\citeauthoryear{{Andredakis}, {Peletier} \&
  {Balcells}}{{Andredakis} et~al.}{1995}]{Andredakis95}
{Andredakis} Y.~C.,  {Peletier} R.~F.,    {Balcells} M.,  1995, \mnras, 275,
  874

\bibitem[\protect\citeauthoryear{{Andredakis} \& {Sanders}}{{Andredakis} \&
  {Sanders}}{1994}]{Andredakis94}
{Andredakis} Y.~C.,  {Sanders} R.~H.,  1994, \mnras, 267, 283

\bibitem[\protect\citeauthoryear{{Baggett}, {Baggett} \& {Anderson}}{{Baggett}
  et~al.}{1998}]{Baggett98}
{Baggett} W.~E.,  {Baggett} S.~M.,    {Anderson} K.~S.~J.,  1998, \aj, 116,
  1626

\bibitem[\protect\citeauthoryear{{Balcells}, {Graham},
  {Dom{\'{\i}}nguez-Palmero} \& {Peletier}}{{Balcells}
  et~al.}{2003}]{Balcells03}
{Balcells} M.,  {Graham} A.~W.,  {Dom{\'{\i}}nguez-Palmero} L.,    {Peletier}
  R.~F.,  2003, \apjl, 582, L79

\bibitem[\protect\citeauthoryear{{Battinelli}, {Capuzzo-Dolcetta}, {Hodge},
  {Vicari} \& {Wyder}}{{Battinelli} et~al.}{2000}]{Battinelli00}
{Battinelli} P.,  {Capuzzo-Dolcetta} R.,  {Hodge} P.~W.,  {Vicari} A.,
  {Wyder} T.~K.,  2000, \aap, 357, 437

\bibitem[\protect\citeauthoryear{{Binney} \& {Tremaine}}{{Binney} \&
  {Tremaine}}{1987}]{Binney87}
{Binney} J.,  {Tremaine} S.,  1987, {Galactic dynamics}.
Princeton, NJ, Princeton University Press, 1987, 747 p.

\bibitem[\protect\citeauthoryear{{Buta}, {van Driel}, {Braine}, {Combes},
  {Wakamatsu}, {Sofue} \& {Tomita}}{{Buta} et~al.}{1995}]{Buta95}
{Buta} R.,  {van Driel} W.,  {Braine} J.,  {Combes} F.,  {Wakamatsu} K.,
  {Sofue} Y.,    {Tomita} A.,  1995, \apj, 450, 593

\bibitem[\protect\citeauthoryear{{Byun} \& {Freeman}}{{Byun} \&
  {Freeman}}{1995}]{Byun95}
{Byun} Y.~I.,  {Freeman} K.~C.,  1995, \apj, 448, 563

\bibitem[\protect\citeauthoryear{{C{\^ o}t{\' e}}, {Carignan} \&
  {Freeman}}{{C{\^ o}t{\' e}} et~al.}{2000}]{Cote00}
{C{\^ o}t{\' e}} S.,  {Carignan} C.,    {Freeman} K.~C.,  2000, \aj, 120, 3027

\bibitem[\protect\citeauthoryear{{Caon}, {Capaccioli} \& {D'Onofrio}}{{Caon}
  et~al.}{1993}]{Caon93}
{Caon} N.,  {Capaccioli} M.,    {D'Onofrio} M.,  1993, \mnras, 265, 1013

\bibitem[\protect\citeauthoryear{{Carignan} \& {Freeman}}{{Carignan} \&
  {Freeman}}{1988}]{Carignan88}
{Carignan} C.,  {Freeman} K.~C.,  1988, \apjl, 332, L33

\bibitem[\protect\citeauthoryear{{Carollo}}{{Carollo}}{1999}]{Carollo99}
{Carollo} C.~M.,  1999, \apj, 523, 566

\bibitem[\protect\citeauthoryear{{Carollo}, {Stiavelli}, {de Zeeuw} \&
  {Mack}}{{Carollo} et~al.}{1997}]{Carollo97}
{Carollo} C.~M.,  {Stiavelli} M.,  {de Zeeuw} P.~T.,    {Mack} J.,  1997, \aj,
  114, 2366

\bibitem[\protect\citeauthoryear{{Carollo}, {Stiavelli} \& {Mack}}{{Carollo}
  et~al.}{1998}]{Carollo98}
{Carollo} C.~M.,  {Stiavelli} M.,    {Mack} J.,  1998, \aj, 116, 68

\bibitem[\protect\citeauthoryear{{Courteau}, {de Jong} \& {Broeils}}{{Courteau}
  et~al.}{1996}]{Courteau96}
{Courteau} S.,  {de Jong} R.~S.,    {Broeils} A.~H.,  1996, \apjl, 457, L73

\bibitem[\protect\citeauthoryear{{de Blok} \& {McGaugh}}{{de Blok} \&
  {McGaugh}}{1997}]{DeBlok97}
{de Blok} W.~J.~G.,  {McGaugh} S.~S.,  1997, \mnras, 290, 533

\bibitem[\protect\citeauthoryear{{de Grijs}}{{de Grijs}}{1998}]{DeGrijs98}
{de Grijs} R.,  1998, \mnras, 299, 595

\bibitem[\protect\citeauthoryear{{de Jong}}{{de Jong}}{1996a}]{DeJong96}
{de Jong} R.~S.,  1996a, \aaps, 118, 557

\bibitem[\protect\citeauthoryear{{de Jong}}{{de Jong}}{1996b}]{DeJong96b}
{de Jong} R.~S.,  1996b, \aap, 313, 377

\bibitem[\protect\citeauthoryear{{de Jong}, {Simard}, {Davies}, {Saglia},
  {Burstein}, {Colless}, {McMahan} \& {Wegner}}{{de Jong}
  et~al.}{2004}]{DeJong04}
{de Jong} R.~S.,  {Simard} L.,  {Davies} R.~L.,  {Saglia} R.~P.,  {Burstein}
  D.,  {Colless} M.,  {McMahan} R.,    {Wegner} G.,  2004, \mnras, 355, 1155

\bibitem[\protect\citeauthoryear{{de Vaucouleurs}}{{de
  Vaucouleurs}}{1948}]{DeVaucouleurs48}
{de Vaucouleurs} G.,  1948, Annales d'Astrophysique, 11, 247

\bibitem[\protect\citeauthoryear{{de Vaucouleurs}}{{de
  Vaucouleurs}}{1958}]{DeVaucouleurs58}
{de Vaucouleurs} G.,  1958, \apj, 128, 465

\bibitem[\protect\citeauthoryear{{de Vaucouleurs}}{{de
  Vaucouleurs}}{1959}]{DeVaucouleurs59}
{de Vaucouleurs} G.,  1959, Handbuch der Physik, 53, 275

\bibitem[\protect\citeauthoryear{{de Vaucouleurs}}{{de
  Vaucouleurs}}{1961}]{DeVaucouleurs61}
{de Vaucouleurs} G.,  1961, \apjs, 5, 233

\bibitem[\protect\citeauthoryear{{Emsellem}, {Cappellari}, {Peletier},
  {McDermid}, {Bacon}, {Bureau}, {Copin}, {Davies}, {Krajnovi{\' c}},
  {Kuntschner}, {Miller} \& {Tim de Zeeuw}}{{Emsellem}
  et~al.}{2004}]{Emsellem04}
{Emsellem} E.,  {Cappellari} M.,  {Peletier} R.~F.,  {McDermid} R.~M.,  {Bacon}
  R.,  {Bureau} M.,  {Copin} Y.,  {Davies} R.~L.,  {Krajnovi{\' c}} D.,
  {Kuntschner} H.,  {Miller} B.~W.,    {Tim de Zeeuw} P.,  2004, \mnras, 352,
  721

\bibitem[\protect\citeauthoryear{{Erwin}, {Beltr{\' a}n}, {Graham} \&
  {Beckman}}{{Erwin} et~al.}{2003}]{Erwin03}
{Erwin} P.,  {Beltr{\' a}n} J.~C.~V.,  {Graham} A.~W.,    {Beckman} J.~E.,
  2003, \apj, 597, 929

\bibitem[\protect\citeauthoryear{{Erwin} \& {Sparke}}{{Erwin} \&
  {Sparke}}{1999}]{Erwin99}
{Erwin} P.,  {Sparke} L.~S.,  1999, \apjl, 521, L37

\bibitem[\protect\citeauthoryear{{Erwin} \& {Sparke}}{{Erwin} \&
  {Sparke}}{2002}]{Erwin02}
{Erwin} P.,  {Sparke} L.~S.,  2002, \aj, 124, 65

\bibitem[\protect\citeauthoryear{{Erwin} \& {Sparke}}{{Erwin} \&
  {Sparke}}{2003}]{Erwin03b}
{Erwin} P.,  {Sparke} L.~S.,  2003, \apjs, 146, 299

\bibitem[\protect\citeauthoryear{Fathi}{Fathi}{2004}]{Fathi04}
Fathi K.,  2004, PhD thesis, Rijksuniversiteit Groningen

\bibitem[\protect\citeauthoryear{{Freeman}}{{Freeman}}{1970}]{Freeman70}
{Freeman} K.~C.,  1970, \apj, 160, 811

\bibitem[\protect\citeauthoryear{{Gonz\'alez Delgado}, {Perez}, {Tadhunter},
  {Vilchez} \& {Rodriguez-Espinosa}}{{Gonz\'alez Delgado}
  et~al.}{1997}]{Delgado97}
{Gonz\'alez Delgado} R.~M.,  {Perez} E.,  {Tadhunter} C.,  {Vilchez} J.~M.,
  {Rodriguez-Espinosa} J.~M.,  1997, \apjs, 108, 155

\bibitem[\protect\citeauthoryear{{Graham}}{{Graham}}{2001}]{Graham01}
{Graham} A.~W.,  2001, \aj, 121, 820

\bibitem[\protect\citeauthoryear{{Huchra}, {Wyatt} \& {Davis}}{{Huchra}
  et~al.}{1982}]{Huchra82}
{Huchra} J.~P.,  {Wyatt} W.~F.,    {Davis} M.,  1982, \aj, 87, 1628

\bibitem[\protect\citeauthoryear{{Hunt}, {Pierini} \& {Giovanardi}}{{Hunt}
  et~al.}{2004}]{Hunt04}
{Hunt} L.~K.,  {Pierini} D.,    {Giovanardi} C.,  2004, \aap, 414, 905

\bibitem[\protect\citeauthoryear{{Illingworth} \& {Schechter}}{{Illingworth} \&
  {Schechter}}{1982}]{Illingworth82}
{Illingworth} G.,  {Schechter} P.~L.,  1982, \apj, 256, 481

\bibitem[\protect\citeauthoryear{{James}, {Shane}, {Beckman}, {Cardwell},
  {Collins}, {Etherton}, {de Jong}, {Fathi}, {Knapen}, {Peletier}, {Percival},
  {Pollacco}, {Seigar}, {Stedman} \& {Steele}}{{James} et~al.}{2004}]{James04}
{James} P.~A.,  {Shane} N.~S.,  {Beckman} J.~E.,  {Cardwell} A.,  {Collins}
  C.~A.,  {Etherton} J.,  {de Jong} R.~S.,  {Fathi} K.,  {Knapen} J.~H.,
  {Peletier} R.~F.,  {Percival} S.~M.,  {Pollacco} D.~L.,  {Seigar} M.~S.,
  {Stedman} S.,    {Steele} I.~A.,  2004, \aap, 414, 23

\bibitem[\protect\citeauthoryear{{Jedrzejewski}}{{Jedrzejewski}}{1987}]{Jedrze%
jewski87}
{Jedrzejewski} R.~I.,  1987, \mnras, 226, 747

\bibitem[\protect\citeauthoryear{{Jerjen}, {Binggeli} \& {Freeman}}{{Jerjen}
  et~al.}{2000}]{Jerjen00}
{Jerjen} H.,  {Binggeli} B.,    {Freeman} K.~C.,  2000, \aj, 119, 593

\bibitem[\protect\citeauthoryear{{Kamphuis}, {Sijbring} \& {van
  Albada}}{{Kamphuis} et~al.}{1996}]{Kamphuis96}
{Kamphuis} J.~J.,  {Sijbring} D.,    {van Albada} T.~S.,  1996, \aaps, 116, 15

\bibitem[\protect\citeauthoryear{{Kennicutt}}{{Kennicutt}}{1989}]{Kennicutt89}
{Kennicutt} R.~C.,  1989, \apj, 344, 685

\bibitem[\protect\citeauthoryear{{Khosroshahi}, {Wadadekar} \&
  {Kembhavi}}{{Khosroshahi} et~al.}{2000}]{Khosroshahi00}
{Khosroshahi} H.~G.,  {Wadadekar} Y.,    {Kembhavi} A.,  2000, \apj, 533, 162

\bibitem[\protect\citeauthoryear{{Kormendy}}{{Kormendy}}{1993}]{Kormendy93}
{Kormendy} J.,  1993, in IAU Symposium 153: Galactic Bulges {Kinematics of
  extragalactic bulges: evidence that some bulges are really disks}.
p.~209

\bibitem[\protect\citeauthoryear{{Kormendy} \& {Illingworth}}{{Kormendy} \&
  {Illingworth}}{1982}]{Kormendy82}
{Kormendy} J.,  {Illingworth} G.,  1982, \apj, 256, 460

\bibitem[\protect\citeauthoryear{{Kormendy} \& {Kennicutt}}{{Kormendy} \&
  {Kennicutt}}{2004}]{Kormendy04}
{Kormendy} J.,  {Kennicutt} R.~C.,  2004, \araa, 42, 603

\bibitem[\protect\citeauthoryear{{Landolt}}{{Landolt}}{1992}]{Landolt92}
{Landolt} A.~U.,  1992, \aj, 104, 340

\bibitem[\protect\citeauthoryear{{Laurikainen}, {Salo} \& {Buta}}{{Laurikainen}
  et~al.}{2005}]{Laurikainen05}
{Laurikainen} E.,  {Salo} H.,    {Buta} R.,  2005, \mnras, 362, 1319

\bibitem[\protect\citeauthoryear{{Laurikainen}, {Salo}, {Buta}, {Knapen},
  {Speltincx} \& {Block}}{{Laurikainen} et~al.}{2006}]{Laurikainen06}
{Laurikainen} E.,  {Salo} H.,  {Buta} R.,  {Knapen} J.,  {Speltincx} T.,
  {Block} D.,  2006, ArXiv Astrophysics e-prints: astro-ph/0609343

\bibitem[\protect\citeauthoryear{{MacArthur}, {Courteau} \&
  {Holtzman}}{{MacArthur} et~al.}{2003}]{MacArthur03}
{MacArthur} L.~A.,  {Courteau} S.,    {Holtzman} J.~A.,  2003, \apj, 582, 689

\bibitem[\protect\citeauthoryear{{Matthews}, {Gallagher} \& {van
  Driel}}{{Matthews} et~al.}{1999}]{Matthews99}
{Matthews} L.~D.,  {Gallagher} J.~S.,    {van Driel} W.,  1999, \aj, 118, 2751

\bibitem[\protect\citeauthoryear{{M{\"o}llenhoff}, {Popescu} \&
  {Tuffs}}{{M{\"o}llenhoff} et~al.}{2006}]{Mollenhoff06}
{M{\"o}llenhoff} C.,  {Popescu} C.~C.,    {Tuffs} R.~J.,  2006, ArXiv
  Astrophysics e-prints: astro-ph/0606562

\bibitem[\protect\citeauthoryear{{Morganti}, {de Zeeuw}, {Oosterloo},
  {McDermid}, {Krajnovic}, {Cappellari}, {Kenn}, {Weijmans} \&
  {Sarzi}}{{Morganti} et~al.}{2006}]{Morganti06}
{Morganti} R.,  {de Zeeuw} P.~T.,  {Oosterloo} T.~A.,  {McDermid} R.~M.,
  {Krajnovic} D.,  {Cappellari} M.,  {Kenn} F.,  {Weijmans} A.,    {Sarzi} M.,
  2006, ArXiv Astrophysics e-prints: astro-ph/0606261

\bibitem[\protect\citeauthoryear{{Nagar}, {Wilson}, {Mulchaey} \&
  {Gallimore}}{{Nagar} et~al.}{1999}]{Nagar99}
{Nagar} N.~M.,  {Wilson} A.~S.,  {Mulchaey} J.~S.,    {Gallimore} J.~F.,  1999,
  \apjs, 120, 209

\bibitem[\protect\citeauthoryear{{Noordermeer}, {van der Hulst}, {Sancisi},
  {Swaters} \& {van Albada}}{{Noordermeer} et~al.}{2005}]{Noordermeer05}
{Noordermeer} E.,  {van der Hulst} J.~M.,  {Sancisi} R.,  {Swaters} R.~A.,
  {van Albada} T.~S.,  2005, \aap, 442, 137 (paper~I)

\bibitem[\protect\citeauthoryear{{Noordermeer}, {van der Hulst}, {Sancisi},
  {Swaters} \& {van Albada}}{{Noordermeer} et~al.}{2006}]{Noordermeer06b}
{Noordermeer} E.,  {van der Hulst} J.~M.,  {Sancisi} R.,  {Swaters} R.~A.,
  {van Albada} T.~S.,  2006, in press

\bibitem[\protect\citeauthoryear{{Oosterloo}, {Morganti}, {Sadler}, {Vergani}
  \& {Caldwell}}{{Oosterloo} et~al.}{2002}]{Oosterloo02}
{Oosterloo} T.~A.,  {Morganti} R.,  {Sadler} E.~M.,  {Vergani} D.,
  {Caldwell} N.,  2002, \aj, 123, 729

\bibitem[\protect\citeauthoryear{{Palunas} \& {Williams}}{{Palunas} \&
  {Williams}}{2000}]{Palunas00}
{Palunas} P.,  {Williams} T.~B.,  2000, \aj, 120, 2884

\bibitem[\protect\citeauthoryear{{Pogge}}{{Pogge}}{1989}]{Pogge89b}
{Pogge} R.~W.,  1989, \apjs, 71, 433

\bibitem[\protect\citeauthoryear{{Rix} \& {White}}{{Rix} \&
  {White}}{1990}]{Rix90}
{Rix} H.-W.,  {White} S.~D.~M.,  1990, \apj, 362, 52

\bibitem[\protect\citeauthoryear{{Roberts} \& {Haynes}}{{Roberts} \&
  {Haynes}}{1994}]{Roberts94}
{Roberts} M.~S.,  {Haynes} M.~P.,  1994, \araa, 32, 115

\bibitem[\protect\citeauthoryear{{Russell}, {Lasker}, {McLean}, {Sturch} \&
  {Jenkner}}{{Russell} et~al.}{1990}]{Russell90}
{Russell} J.~L.,  {Lasker} B.~M.,  {McLean} B.~J.,  {Sturch} C.~R.,
  {Jenkner} H.,  1990, \aj, 99, 2059

\bibitem[\protect\citeauthoryear{{Schlegel}, {Finkbeiner} \&
  {Davis}}{{Schlegel} et~al.}{1998}]{Schlegel98}
{Schlegel} D.~J.,  {Finkbeiner} D.~P.,    {Davis} M.,  1998, \apj, 500, 525

\bibitem[\protect\citeauthoryear{{Scorza} \& {Bender}}{{Scorza} \&
  {Bender}}{1995}]{Scorza95}
{Scorza} C.,  {Bender} R.,  1995, \aap, 293, 20

\bibitem[\protect\citeauthoryear{{S\'ersic}}{{S\'ersic}}{1968}]{Sersic68}
{S\'ersic} J.~L.,  1968, {Atlas de Galaxias Australes}.
Cordoba, Argentina: Observatorio Astronomico, 1968

\bibitem[\protect\citeauthoryear{Swaters}{Swaters}{1999}]{Swatersthesis}
Swaters R.,  1999, PhD thesis, Rijksuniversiteit Groningen

\bibitem[\protect\citeauthoryear{{Trujillo}, {Aguerri}, {Cepa} \&
  {Guti{\'e}rrez}}{{Trujillo} et~al.}{2001}]{Trujillo01b}
{Trujillo} I.,  {Aguerri} J.~A.~L.,  {Cepa} J.,    {Guti{\'e}rrez} C.~M.,
  2001, \mnras, 321, 269

\bibitem[\protect\citeauthoryear{{Trujillo}, {Graham} \& {Caon}}{{Trujillo}
  et~al.}{2001}]{Trujillo01}
{Trujillo} I.,  {Graham} A.~W.,    {Caon} N.,  2001, \mnras, 326, 869

\bibitem[\protect\citeauthoryear{{Ulvestad} \& {Wilson}}{{Ulvestad} \&
  {Wilson}}{1984}]{Ulvestad84}
{Ulvestad} J.~S.,  {Wilson} A.~S.,  1984, \apj, 285, 439

\bibitem[\protect\citeauthoryear{{van der Hulst}, {van Albada} \&
  {Sancisi}}{{van der Hulst} et~al.}{2001}]{VanderHulst01}
{van der Hulst} J.~M.,  {van Albada} T.~S.,    {Sancisi} R.,  2001, in ASP
  Conf. Ser. 240: Gas and Galaxy Evolution {The Westerbork HI Survey of
  Irregular and Spiral Galaxies, WHISP}.
p.~451

\bibitem[\protect\citeauthoryear{{van der Kruit}}{{van der
  Kruit}}{1979}]{VanderKruit79}
{van der Kruit} P.~C.,  1979, \aaps, 38, 15

\bibitem[\protect\citeauthoryear{{van Driel} \& {Buta}}{{van Driel} \&
  {Buta}}{1991}]{VanDriel91b}
{van Driel} W.,  {Buta} R.~J.,  1991, \aap, 245, 7

\bibitem[\protect\citeauthoryear{{Verdes-Montenegro}, {Bosma} \&
  {Athanassoula}}{{Verdes-Montenegro} et~al.}{1995}]{Verdes-Montenegro95}
{Verdes-Montenegro} L.,  {Bosma} A.,    {Athanassoula} E.,  1995, \aap, 300, 65

\bibitem[\protect\citeauthoryear{{V\'eron-Cetty} \& {V\'eron}}{{V\'eron-Cetty}
  \& {V\'eron}}{1996}]{Veron-Cetty96}
{V\'eron-Cetty} M.-P.,  {V\'eron} P.,  1996, \aaps, 115, 97

\bibitem[\protect\citeauthoryear{{Young} \& {Currie}}{{Young} \&
  {Currie}}{1994}]{Young94}
{Young} C.~K.,  {Currie} M.~J.,  1994, \mnras, 268, L11

\bibitem[\protect\citeauthoryear{{Zwicky} \& {Zwicky}}{{Zwicky} \&
  {Zwicky}}{1971}]{Zwicky71}
{Zwicky} F.,  {Zwicky} M.~A.,  1971, {Catalogue of selected compact galaxies
  and of post-eruptive galaxies}.
Guemligen: Zwicky, 1971

\end{thebibliography}

\appendix

\section{Tables}
\label{app:tables}

\begin{table*}
 \begin{minipage}{13.cm}
  \centering
  \caption[List of photometric observations]
  {List of observations: (1)~galaxy name; (2)~colour band;
   (3)~telescope used; (4)~observing dates; (5)~total exposure time; 
   (6)~effective seeing; (7)~photometric error and (8)~magnitude  
   corresponding to $3\sigma_{\mathrm {bg}}$ above sky level. See text
   for detailed description of the
   columns.\label{table:observations}}  
  
  \begin{tabular}{rcclrclcc}
    \hline 
    \multicolumn{1}{c}{UGC} & band & telescope &
    \multicolumn{1}{c}{dates} & \multicolumn{1}{c}{t$_{\mathrm 
    {exp}}$} & seeing & \multicolumn{1}{c}{$\sigma_{\mathrm {phot}}$} & 
    \multicolumn{1}{c}{$\mu_{3\sigma}$} & \\  
  
    & & & & \multicolumn{1}{c}{sec} & \arcsec & \multicolumn{1}{c}{mag}
    & \multicolumn{1}{c}{mag} & \\
    
    \multicolumn{1}{c}{(1)} & (2) & (3) & \multicolumn{1}{c}{(4)} &
    \multicolumn{1}{c}{(5)} & (6) & \multicolumn{1}{c}{(7)} &
    \multicolumn{1}{c}{(8)} & \\

    \hline 
    89   & R                           & JKT & 2/11/00$^\ddagger$, 4/11/00  & 2800      & 1.1 & 0.076        & 26.12 & \\
         & B                           & JKT & 4/11/00                      & 2400      & 1.3 & 0.12         & 26.63 & \\ 
    94   & R                           & INT & 5/11/00                      & 600       & 1.5 & 0.038        & 25.77 & \\ 
         & B                           & INT & 5/11/00                      & 600       & 1.7 & 0.029        & 27.77 & \\ 
    624  & R                           & JKT & 31/10/00                     & 2400      & 0.8 & 0.049        & 26.71 & \\ 
         & B                           & INT & 5/11/00                      & 600       & 1.8 & 0.030        & 28.64 & \\ 
         & I                           & JKT & 27/1/01                      & 1440      & 1.3 & 0.049        & 24.19 & \\ 
    1541 & R                           & JKT & 2/11/00$^\ddagger$, 3/11/00  & 2880      & 1.6 & 0.093        & 26.23 & \\ 
         & B                           & JKT & 3/11/00                      & 2400      & 2.2 & 0.16         & 27.72 & \\ 
         & I                           & JKT & 27/1/01                      & 1500      & 1.4 & 0.047        & 24.48 & \\ 
    2487 & R                           & JKT & 29/10/00, 30/10/00           & 2400      & 1.8 & 0.027        & 26.36 & \\ 
         & B                           & JKT & 24/1/01                      & 2400      & 1.3 & 0.072        & 26.46 & \\ 
    2916 & R                           & JKT & 31/10/00                     & 2400      & 1.0 & 0.056        & 26.80 & \\ 
         & B                           & JKT & 5/11/00                      & 2400      & 1.8 & 0.019        & 27.38 & \\ 
    2953 & R$^\dagger$\hspace{-0.14cm} & JKT & 18/11/01$^\ddagger$          & 2400      & 1.7 & 0.029$^\$$   & 24.83 & \\ 
         & R$^\dagger$\hspace{-0.14cm} & INT & 3/2/95$^\ddagger$            &  180      & 2.2 & 0.51$^\$$    & 26.83 & \\
         & B$^\dagger$\hspace{-0.14cm} & JKT & 4/11/00, 5/11/00             & 3300      & 1.9 & 0.14         & 25.47 & \\ 
         & B$^\dagger$\hspace{-0.14cm} & JKT & 18/11/01$^\ddagger$          & 2400      & 2.1 & 0.055$^\$$   & 25.43 & \\ 
         & I                           & JKT & 29/1/01                      & 1800      & 1.1 & 0.15         & 23.05 & \\ 
    3205 & R$^\dagger$\hspace{-0.14cm} & JKT & 26/1/01                      & 1800      & 1.3 & 0.12         & 25.59 & \\ 
         & R$^\dagger$\hspace{-0.14cm} & JKT & 29/10/00                     & 2400      & 3.6 & 0.032        & 25.22 & \\ 
         & B                           & JKT & 1/11/00                      & 2400      & 1.0 & 0.17         & 26.68 & \\ 
         & I                           & JKT & 30/1/01                      & 2000      & 3.1 & 0.20         & 23.92 & \\ 
    3546 & R$^\dagger$\hspace{-0.14cm} & JKT & 30/10/00                     & 2400      & 0.9 & 0.028        & 26.27 & \\ 
         & R$^\dagger$\hspace{-0.14cm} & JKT & 26/1/01                      & 1800      & 1.2 & 0.13         & 26.03 & \\ 
         & R$^\dagger$\hspace{-0.14cm} & JKT & 18/11/01$^\ddagger$          & 2400      & 2.0 & 0.028$^\$$   & 26.79 & \\ 
         & B$^\dagger$\hspace{-0.14cm} & JKT & 30/10/00, 5/11/00            & 2400      & 1.6 & 0.22         & 27.77 & \\ 
         & B$^\dagger$\hspace{-0.14cm} & JKT & 18/11/01$^\ddagger$          & 2400      & 2.0 & 0.052$^\$$   & 27.09 & \\ 
    3580 & R                           & INT & 27/12/95                     &  300      & 1.0 & 0.051        & 26.20 & \\
         & B                           & INT & 27/12/95                     &  300      & 1.2 & 0.12         & 28.05 & \\
    3993 & R                           & JKT & 26/1/01                      & 2400      & 1.2 & 0.16         & 26.86 & \\ 
         & B                           & JKT & 2/4/02$^*$                   & 500       & 1.7 & 0.18         & 25.74 & \\ 
    4458 & R                           & JKT & 2/11/00$^\ddagger$, 3/11/00  & 3680      & 1.1 & 0.095        & 26.56 & \\
         & B$^\dagger$\hspace{-0.14cm} & JKT & 25/1/01                      & 2400      & 2.0 & 0.18         & 27.15 & \\
         & B$^\dagger$\hspace{-0.14cm} & JKT & 3/11/00                      & 2400      & 2.1 & 0.15         & 26.32 & \\
    5253 & R                           & INT & 3/5/94                       & 300       & 1.6 & 0.24         & 26.84 & \\
         & B                           & JKT & 4/4/02$^{*\ddagger}$         & 1800      & 1.6 & 0.14$^\$$    & 26.19 & \\
    6786 & R$^\dagger$\hspace{-0.14cm} & JKT & 31/1/01                      & 720       & 1.1 & 0.053        & 25.69 & \\ 
         & R$^\dagger$\hspace{-0.14cm} & MDM & 1/1/03                       & 480       & 1.4 & 0.022        & 25.63 & \\ 
    6787 & R                           & JKT & 29/5/00                      & 1800      & 1.7 & 0.26         & 26.41 & \\ 
         & B                           & JKT & 31/1/01                      & 1800      & 1.2 & 0.11         & 27.82 & \\ 
    8699 & R                           & JKT & 24/5/00                      & 2400      & 1.6 & 0.24         & 25.33 & \\
         & B                           & JKT & 24/5/00                      & 2400      & 1.7 & 0.11         & 25.76 & \\
    9133 & R                           & JKT & 26/5/00, 27/5/00             & 2400      & 1.3 & 0.086        & 25.61 & \\ 
         & B                           & JKT & 1/6/00, 19/4/02$^*$          & 3000      & 2.3 & 0.085        & 27.23 & \\ 
    11670& R                           & JKT & 26/5/00                      & 3000      & 1.3 & 0.11         & 25.98 & \\ 
         & B                           & JKT & 30/5/00$^\ddagger$           & 2400      & 1.3 & 0.19$^\$$    & 27.01 & \\ 
         & I                           & JKT & 31/5/00$^\ddagger$, 1/6/00   & 2400      & 1.1 & 0.14         & 25.12 & \\ 
    11852& R                           & JKT & 27/5/00                      & 2400      & 0.9 & 0.081        & 25.86 & \\  
         & B                           & JKT & 27/5/00                      & 2400      & 1.1 & 0.084        & 26.58 & \\ 
         & I                           & JKT & 1/6/00                       & 2400      & 1.1 & 0.075        & 24.82 & \\ 
    11914& R                           & JKT & 30/5/00$^\ddagger$, 31/5/00$^\ddagger$, 3/6/02$^*$ & 2900 & 1.1 & 0.22   & 25.40 & \\
         & B                           & JKT &30/5/00$^\ddagger$, 3/6/02$^*$& 2900      & 1.3 & 0.24         & 26.65 & \\ 
    12043& R                           & JKT & 29/5/00                      & 2400      & 1.2 & 0.26         & 26.04 & \\
         & B                           & JKT & 31/5/00$^\ddagger$, 1/6/01   & 3300      & 1.4 & 0.12         & 26.37 & \\
    \hline
    \multicolumn{9}{l}{$^\dagger$ multiple exposures used to check consistency
                       of our observations, see section~\ref{sec:comparison}.} \\  
    \multicolumn{9}{l}{\hspace{0.14cm} Only the first observation listed for
                       each band is used in the remainder of this paper.} \\    
    \multicolumn{9}{l}{$^\ddagger$ night with non-photometric conditions.} \\
    \multicolumn{9}{l}{$^\$$ lower limit for photometric error only, due to
                       non-photometric conditions during observations.} \\ 
    \multicolumn{8}{l}{$^*$ observations done in service mode at the JKT.} \\ 
    \multicolumn{8}{l}{$^\#$mosaic of 4 pointings.} \\
  \end{tabular}
 \end{minipage}
\end{table*}  

\begin{table*}
 \begin{minipage}{11.1cm}
  \centering
  \caption[Isophotal properties]
  {Isophotal properties: (1)~UGC number; (2)~RA and (3)~Dec of central 
   position; (4)~uncertainty in central position and (5)~dominant 
   source of uncertainty; (6)~position angle (north through east) of
   major axis; (7)~ellipticity and (8)~inclination
   angle. \label{table:isoresults}}  
  
  \begin{tabular}{rr@{\hspace{0.2cm}}r@{\hspace{0.25cm}}d{2.2}@{\hspace{0.7cm}}r@{\hspace{0.25cm}}r@{\hspace{0.cm}}r@{\hspace{0.7cm}}ccrrr}   
    \hline 
    \multicolumn{1}{c}{UGC} & \multicolumn{6}{c}{central position} & 
    \multicolumn{2}{c}{uncertainty} & \multicolumn{1}{c}{PA} &
    \multicolumn{1}{c}{$\epsilon$} & \multicolumn{1}{c}{i} \\   

    & \multicolumn{3}{c@{\hspace{0.7cm}}}{RA (2000)} &
    \multicolumn{3}{c@{\hspace{0.9cm}}}{Dec (2000)} &  
    $\sigma_{\mathrm {pos}}$ & source & & & \\

    & \multicolumn{1}{c}{\it h} &
    \multicolumn{1}{c@{\hspace{0.3cm}}}{\it m} & 
    \multicolumn{1}{c@{\hspace{0.9cm}}}{\it s} &
    \multicolumn{1}{c@{\hspace{0.35cm}}}{$^{\circ}$} &
    \multicolumn{1}{c@{\hspace{0.1cm}}}{\arcmin} & 
    \multicolumn{1}{c@{\hspace{0.6cm}}}{\arcsec} & \arcsec & &
    \multicolumn{1}{c}{$^{\circ}$} & & \multicolumn{1}{c}{$^{\circ}$} \\     
    
    \multicolumn{1}{c}{(1)} &
    \multicolumn{3}{c@{\hspace{0.7cm}}}{(2)} & 
    \multicolumn{3}{c@{\hspace{0.7cm}}}{(3)} & (4) & (5) &
    \multicolumn{1}{c}{(6)} & \multicolumn{1}{c}{(7)} &
    \multicolumn{1}{c}{(8)} \\   
    \hline
    89    &  0 &  9 & 53.47 & 25 & 55 & 26.0 & 0.6 & DSS  & -14 & 0.37 & 52 \\
    94    &  0 & 10 & 26.05 & 25 & 49 & 47.5 & 0.6 & DSS  & -76 & 0.34 & 50 \\
    624   &  1 &  0 & 36.42 & 30 & 40 &  8.6 & 0.6 & DSS  & -72 & 0.52 & 64 \\
    1541  &  2 &  3 & 27.97 & 38 &  7 &  0.9 & 0.6 & DSS  &  67 & 0.26 & 43 \\
    2487  &  3 &  1 & 42.34 & 35 & 12 & 20.6 & 0.6 & DSS  &  70 & 0.20 & 38 \\
    2916  &  4 &  2 & 33.8  & 71 & 42 & 21.1 & 0.6 & DSS  &  53 & 0.14 & 31 \\
    2953  &  4 &  7 & 46.7  & 69 & 48 & 46.7 & 0.6 & DSS  & -85 & 0.33 & 49 \\
    3205  &  4 & 56 & 14.88 & 30 &  3 &  8.3 & 0.6 & DSS  &  42 & 0.60 & 69 \\
    3546  &  6 & 50 &  8.7  & 60 & 50 & 45.4 & 0.6 & DSS  &  51 & 0.37 & 52 \\
    3580  &  6 & 55 & 30.9  & 69 & 33 & 47.1 & 1.0 & dust &   7 & 0.48 & 61 \\
    3993  &  7 & 55 & 44.1  & 84 & 55 & 35.4 & 0.6 & DSS  &  44 & 0.08 & 23 \\
    4458  &  8 & 32 & 11.34 & 22 & 33 & 38.3 & 0.6 & DSS  & -72 & 0.12 & 29 \\
    5253  &  9 & 50 & 22.50 & 72 & 16 & 44.3 & 0.6 & DSS  &  -3 & 0.18 & 36 \\
    6786  & 11 & 49 &  9.52 & 27 &  1 & 20.2 & 0.6 & DSS  &   1 & 0.45 & 58$^{\dagger}$\hspace{-0.14cm} \\
    6787  & 11 & 49 & 15.26 & 56 &  5 &  5.5 & 0.6 & DSS  & -73 & 0.40 & 55$^{\dagger}$\hspace{-0.14cm} \\
    8699  & 13 & 45 &  8.74 & 41 & 30 & 12.3 & 0.6 & DSS  & -81 & 0.70 & 77 \\
    9133  & 14 & 16 &  7.76 & 35 & 20 & 38.3 & 0.6 & DSS  &  26 & 0.37 & 52 \\
    11670 & 21 &  3 & 33.56 & 29 & 53 & 51.2 & 0.6 & DSS  & -26 & 0.58 & 68 \\
    11852 & 21 & 55 & 59.27 & 27 & 53 & 55.6 & 0.6 & DSS  &  13 & 0.38 & 53 \\
    11914 & 22 &  7 & 52.43 & 31 & 21 & 34.5 & 0.6 & DSS  &  81 & 0.10 & 26 \\
    12043 & 22 & 27 & 50.53 & 29 &  5 & 45.9 & 0.7 & DSS  & -84 & 0.60 & 69 \\
    \hline 
    \multicolumn{12}{l}{$^\dagger$But isophotes not representative of
                        disk ellipticity; see text.} 
  \end{tabular}
 \end{minipage}
\end{table*}  

\begin{table*}
 \begin{minipage}{16.9cm}
  \centering
  \caption[Photometric properties]
  {Photometric properties: (1)~UGC number; (2)~colour band for columns 
   (3)~to (14); (3)~and (4)~raw isophotal diameters at 25 and 26.5
   \magasas; (5)~and (6)~idem, but corrected for inclination and
   Galactic foreground extinction; (7) -- (9)~effective radii
   containing 20, 50 and 80\% of the total light; (10)~conversion
   factor to convert arcseconds into kpc; (11)~and (12)~total apparent 
   magnitudes within the 25th\magasas\ diameter and within the last
   measured point; (13)~and (14)~total absolute magnitudes (corrected for
   Galactic foreground extinction).
  \label{table:photresults}}  

  \begin{tabular}{rcrrrrrrrd{1.2}r@{\hspace{0.1cm}$\pm$\hspace{0.1cm}}rr@{\hspace{0.1cm}$\pm$\hspace{0.1cm}}rrr}  
    \hline 
    \multicolumn{1}{c}{UGC} & band & \multicolumn{1}{c}{$D_{25}$} &
    \multicolumn{1}{c}{$D_{26.5}$} & \multicolumn{1}{c}{$D^{c}_{25}$}
    & \multicolumn{1}{c}{$D^{c}_{26.5}$} &
    \multicolumn{1}{c}{$R_{20}$} & \multicolumn{1}{c}{$R_{50}$} &
    \multicolumn{1}{c}{$R_{80}$} & \multicolumn{1}{c}{scale} &
    \multicolumn{2}{c}{$m_{25}$} & \multicolumn{2}{c}{$m_{\mathrm{lim}}$} &
    \multicolumn{1}{c}{$M_{25}$} & \multicolumn{1}{c}{$M_{\mathrm{lim}}$} \\  

    & & \multicolumn{1}{c}{\arcsec} & \multicolumn{1}{c}{\arcsec} & 
    \multicolumn{1}{c}{\arcsec} & \multicolumn{1}{c}{\arcsec} & 
    \multicolumn{1}{c}{\arcsec} & \multicolumn{1}{c}{\arcsec} &
    \multicolumn{1}{c}{\arcsec} & \multicolumn{1}{c}{kpc/\arcsec} &
    \multicolumn{2}{c}{mag} & \multicolumn{2}{c}{mag} &
    \multicolumn{1}{c}{mag} & \multicolumn{1}{c}{mag} \\   
      
    \multicolumn{1}{c}{(1)} & (2) & 
    \multicolumn{1}{c}{(3)} & \multicolumn{1}{c}{(4)} &
    \multicolumn{1}{c}{(5)} & \multicolumn{1}{c}{(6)} &
    \multicolumn{1}{c}{(7)} & \multicolumn{1}{c}{(8)} &
    \multicolumn{1}{c}{(9)} & \multicolumn{1}{c}{(10)} &
    \multicolumn{2}{c}{(11)} & \multicolumn{2}{c}{(12)}  &
    \multicolumn{1}{c}{(13)} & \multicolumn{1}{c}{(14)} \\  
    \hline 
    89    & R & 197 & --  & 181 & 237 & 4.4 & 23 & 40 & 0.30  & 11.41 & 0.08           & 11.38 & 0.08 & -22.66 & -22.69\\
          & B & 145 & 202 & 129 & 191 & 4.2 & 23 & 40 & 0.30  & 12.76 & 0.12           & 12.68 & 0.13 & -21.37 & -21.45\\
    94    & R & 107 & --  & 98  & 136 & 9.8 & 12 & 28 & 0.30  & 12.72 & 0.04           & 12.69 & 0.04 & -21.36 & -21.39\\
          & B & 87  & 121 & 82  & 107 & 9.9 & 15 & 31 & 0.30  & 13.92 & 0.03           & 13.83 & 0.03 & -20.23 & -20.32\\
    624   & R & 192 & 258 & 145 & 225 & 5.3 & 18 & 42 & 0.32  & 11.98 & 0.05           & 11.96 & 0.05 & -22.23 & -22.25\\
          & B & 112 & 191 & 91  & 149 & 6.6 & 20 & 42 & 0.32  & 13.61 & 0.03           & 13.48 & 0.03 & -20.70 & -20.83\\
          & I & --  & --  & 166 & --  & 5.0 & 16 & 37 & 0.32  & \multicolumn{2}{c}{--} & 11.23 & 0.07 &  --    & -22.94\\
    1541  & R & 145 & --  & 137 & --  & 4.0 & 19 & 39 & 0.37  & 12.21 & 0.09           & 12.17 & 0.09 & -22.37 & -22.41\\
          & B & 106 & 167 & 104 & 163 & 5.3 & 27 & 44 & 0.37  & 13.73 & 0.16           & 13.55 & 0.16 & -20.94 & -21.12\\
          & I & --  & --  & --  & --  & 3.9 & 16 & 37 & 0.37  & \multicolumn{2}{c}{--} & 11.47 & 0.05 &  --    & -23.07\\
    2487  & R & 211 & 275 & 220 & --  & 9.3 & 30 & 63 & 0.33  & 11.43 & 0.03           & 11.39 & 0.03 & -23.20 & -23.24\\
          & B & 143 & 217 & 166 & --  & 9.8 & 32 & 68 & 0.33  & 13.27 & 0.08           & 13.06 & 0.10 & -21.67 & -21.88\\
    2916  & R & 113 & 148 & 125 & --  & 5.0 & 14 & 33 & 0.31  & 12.80 & 0.06           & 12.75 & 0.06 & -21.96 & -22.01\\
          & B & 80  & 118 & 106 & --  & 6.3 & 17 & 35 & 0.31  & 14.34 & 0.02           & 14.18 & 0.03 & -20.89 & -21.05\\
    2953  & R & 531$^{\$}$\hspace{-0.14cm} & --  & --  & --  & 32 & 81 & 150 & 0.073 & 9.49 & 0.05$^{\$}$\hspace{-0.15cm} & 9.48 & 0.05$^{\$}$\hspace{-0.14cm} & -22.53 & -22.54\\
          & B & 338 & --  & --  & --  & 32  & 77 & 133& 0.073 & 11.56 & 0.15           & 11.49 & 0.16 & -21.15 & -21.22\\
          & I & --  & --  & --  & --  & 31  & 76 & 144& 0.073 & \multicolumn{2}{c}{--} & 8.34  & 0.16 &  --    & -23.37\\ 
    3205  & R & 142 & --  & 148 & --  & 9.9 & 23 & 40 & 0.24  & 13.02 & 0.12           & 13.00 & 0.12 & -21.86 & -21.88\\
          & B & 90  & 146 & 138 & --  & 12  & 27 & 47 & 0.24  & 15.15 & 0.17           & 14.88 & 0.18 & -20.62 & -20.89\\
          & I & --  & --  & --  & --  & 11  & 24 & 43 & 0.24  & \multicolumn{2}{c}{--} & 12.01 & 0.21 &  --    & -22.48\\
    3546  & R & 234 & --  & 224 & 270 & 8.9 & 25 & 66 & 0.13  & 11.03 & 0.03           & 11.02 & 0.03 & -21.34 & -21.35\\
          & B & 193 & 236 & 185 & 230 & 11  & 28 & 70 & 0.13  & 12.54 & 0.22           & 12.47 & 0.22 & -19.95 & -20.02\\
    3580  & R & 196 & 253 & 164 & 231 & 8.9 & 25 & 58 & 0.093 & 12.18 & 0.04           & 12.14 & 0.05 & -19.38 & -19.42\\
          & B & 140 & 215 & 111 & 189 & 8.6 & 26 & 62 & 0.093 & 13.52 & 0.01           & 13.33 & 0.02 & -18.12 & -18.31\\
    3993  & R & 106 & 135 & 108 & 137 & 3.5 & 11 & 28 & 0.30  & 12.82 & 0.16           & 12.78 & 0.16 & -21.31 & -21.35\\
          & B & 88  & --  & 92  & --  & 4.4 & 14 & 36 & 0.30  & 14.18 & 0.19           & 14.04 & 0.20 & -20.05 & -20.19\\
    4458  & R & 188 & 272 & 186 & 270 & 4.5 & 19 & 56 & 0.31  & 11.58 & 0.10           & 11.52 & 0.10 & -22.55 & -22.61\\
          & B & 134 & 222 & 134 & 222 & 5.6 & 25 & 67 & 0.31  & 13.05 & 0.18           & 12.81 & 0.18 & -21.14 & -21.38\\
    5253  & R & 433 & --  & 417 & 543 & 12  & 42 & 103& 0.10  & 9.81  & 0.24           & 9.79  & 0.24 & -21.88 & -21.90\\
          & B & 354$^{\$}$\hspace{-0.14cm} & --  & 333$^{\$}$\hspace{-0.15cm} & 490$^{\$}$\hspace{-0.14cm} & 17 & 55 & 123& 0.10 & 10.97 & 0.14$^{\$}$\hspace{-0.15cm} & 10.88 & 0.15$^{\$}$\hspace{-0.14cm} & -20.77 & -20.86\\
    6786  & R & 262 & --  & 224 & 323 & 9.6 & 29 & 64 & 0.13  & 11.05 & 0.05           & 11.02 & 0.06 & -21.10 & -21.13\\
    6787  & R & 316 & --  & 288 & 365 & 7.4 & 29 & 75 & 0.092 & 10.17 & 0.26           & 10.16 & 0.26 & -21.27 & -21.28\\
          & B & 260 & 383 & 223 & 340 & 9.5 & 37 & 90 & 0.092 & 11.57 & 0.11           & 11.47 & 0.11 & -19.90 & -20.00\\
    8699  & R & 184 & --  & 118 & 179 & 5.8 & 18 & 42 & 0.18  & 12.14 & 0.24           & 12.11 & 0.24 & -20.71 & -20.74\\
          & B & 137 & --  & 93  & 133 & 6.7 & 21 & 43 & 0.18  & 13.45 & 0.11           & 13.39 & 0.11 & -19.42 & -19.48\\
    9133  & R & 276 & --  & 244 & 336 & 8.6 & 31 & 77 & 0.26  & 11.13 & 0.09           & 11.09 & 0.09 & -22.58 & -22.62\\
          & B & 190 & 274 & 159 & 251 & 11  & 36 & 78 & 0.26  & 12.66 & 0.09           & 12.52 & 0.09 & -21.08 & -21.22\\
    11670 & R & 306 & --  & 282 & 365 & 12  & 34 & 76 & 0.062 & 10.57 & 0.11           & 10.55 & 0.11 & -20.53 & -20.55\\
          & B & 224$^{\$}$\hspace{-0.14cm} & 305$^{\$}$\hspace{-0.14cm} & 218$^{\$}$\hspace{-0.15cm} & 299$^{\$}$\hspace{-0.14cm} & 14  & 36 & 79 & 0.062 & 12.33 & 0.19$^{\$}$\hspace{-0.145cm} & 12.25 & 0.19$^{\$}$\hspace{-0.145cm} & -19.12 & -19.20\\
          & I & 367 & --  & 333 & --  & 13  & 37 & 81 & 0.062 & 9.82  & 0.14           & 9.81  & 0.14 & -21.12 & -21.13\\
    11852 & R & 101 & --  & 93  & --  & 4.8 & 13 & 29 & 0.39  & 13.20 & 0.08           & 13.17 & 0.08 & -21.50 & -21.53\\
          & B & 81  & 121 & 76  & 112 & 6.4 & 16 & 37 & 0.39  & 14.57 & 0.09           & 14.38 & 0.10 & -20.25 & -20.44\\
          & I & --  & --  & 122 & --  & 5.7 & 16 & 37 & 0.39  & \multicolumn{2}{c}{--} & 12.51 & 0.09 &  --    & -22.14\\
    11914 & R & 367 & --  & 374 & --  & 16  & 45 & 91 & 0.072 & 9.77  & 0.22           & 9.75  & 0.22 & -21.33 & -21.35\\
          & B & 251 & 342 & 266 & 356 & 16  & 42 & 81 & 0.072 & 11.04 & 0.24           & 10.98 & 0.24 & -20.21 & -20.27\\
    12043 & R & 129 & --  & 104 & 143 & 8.3 & 18 & 34 & 0.075 & 12.88 & 0.26           & 12.85 & 0.26 & -18.23 & -18.26\\
          & B & 109 & 154 & 89  & 128 & 8.3 & 18 & 35 & 0.075 & 13.75 & 0.12           & 13.68 & 0.12 & -17.46 & -17.53\\
    \hline
    \multicolumn{16}{l}{$^\$$ Values from non-photometric data. Isophotal
    diameters uncertain; errors on magnitudes are lower limits only.}
  \end{tabular} 
 \end{minipage}
\end{table*}  

\begin{table*}
 \begin{minipage}{21cm}
 \vspace{2cm}
 \centering
  \rotcaption[Results from the bulge-disk decompositions]
  {Results from the bulge-disk decompositions: (1)~UGC number;
   (2)~colour band for columns (3)~to (17); (3)~conversion factor to
   convert arcseconds into kpc; (4)~bulge ellipticity; (5)~bulge
   intrinsic axis ratio; (6)~bulge effective surface brightness;
   (7)~idem, but corrected for Galactic foreground extinction;
   (8)~bulge effective radius; (9)~bulge S\'ersic parameter,
   (10)~bulge total apparent magnitude; (11)~bulge total absolute
   magnitude (corrected for Galactic foreground extinction); (12)~disk 
   central surface brightness; (13)~idem, but corrected for inclination 
   and Galactic foreground extinction; (14)~disk scale length; (15)~disk 
   total apparent magnitude; (16)~disk total absolute magnitude 
   (corrected for Galactic foreground extinction) and (17)~total 
   bulge-to-disk luminosity ratio.   
  \label{table:BDresults}} \hspace{-1.35cm}
  \begin{rotate}{90}
   \centering
    \hspace{0cm}%
    \begin{tabular}{rcd{1.2}@{\hspace{0.7cm}}rrrrrrrr@{\hspace{0.7cm}}rrrrr@{\hspace{0.7cm}}r}  
     \hline 
     \multicolumn{1}{c}{UGC} & band &
     \multicolumn{1}{c@{\hspace{0.7cm}}}{scale} & 
     \multicolumn{8}{c@{\hspace{0.7cm}}}{bulge parameters} &  
     \multicolumn{5}{c@{\hspace{0.7cm}}}{disk parameters} & \\

      & & & \multicolumn{1}{c}{$\epsilon_b$} &
     \multicolumn{1}{c}{$q_b$} & \multicolumn{1}{c}{$\mu_e$} &
     \multicolumn{1}{c}{$\mu_e^c$} & \multicolumn{1}{c}{$r_e$} &
     \multicolumn{1}{c}{$n$} & \multicolumn{1}{c}{$m_b$} & 
     \multicolumn{1}{c@{\hspace{0.7cm}}}{$M_b$} &
     \multicolumn{1}{c}{$\mu_0$} & \multicolumn{1}{c}{$\mu_0^c$} &
     \multicolumn{1}{c}{$h$} & \multicolumn{1}{c}{$m_d$} & 
     \multicolumn{1}{c@{\hspace{0.7cm}}}{$M_d$} &
     \multicolumn{1}{c}{B/D} \\ 

      & & \multicolumn{1}{c@{\hspace{0.7cm}}}{kpc/\arcsec} & & & 
     \multicolumn{1}{c}{$\frac{\mathrm{mag}}{\mathrm{arcsec}^2}$} & 
     \multicolumn{1}{c}{$\frac{\mathrm{mag}}{\mathrm{arcsec}^2}$} & 
     \multicolumn{1}{c}{\arcsec} & & \multicolumn{1}{c}{mag} & 
     \multicolumn{1}{c@{\hspace{0.7cm}}}{mag} &   
     \multicolumn{1}{c}{$\frac{\mathrm{mag}}{\mathrm{arcsec}^2}$} &
     \multicolumn{1}{c}{$\frac{\mathrm{mag}}{\mathrm{arcsec}^2}$} &
     \multicolumn{1}{c}{\arcsec} &  
     \multicolumn{1}{c}{mag} &
     \multicolumn{1}{c@{\hspace{0.7cm}}}{mag} & \\

     \multicolumn{1}{c}{(1)} & (2) & \multicolumn{1}{c@{\hspace{0.7cm}}}{(3)} &
     \multicolumn{1}{c}{(4)} & \multicolumn{1}{c}{(5)} &
     \multicolumn{1}{c}{(6)} & \multicolumn{1}{c}{(7)} &
     \multicolumn{1}{c}{(8)} & \multicolumn{1}{c}{(9)} &
     \multicolumn{1}{c}{(10)} & 
     \multicolumn{1}{c@{\hspace{0.7cm}}}{(11)} &
     \multicolumn{1}{c}{(12)} & \multicolumn{1}{c}{(13)} &
     \multicolumn{1}{c}{(14)} & \multicolumn{1}{c}{(15)} & 
     \multicolumn{1}{c@{\hspace{0.7cm}}}{(16)} &
     \multicolumn{1}{c}{(17)} \\    
     \hline 
     89 & R & 0.30 & 0.25 & 0.5 &                   17.53 & 17.43 & 2.9  & 1.2                   & 12.77 & -21.30 & 20.19 & 20.61 & 22.1 & 11.75 & -22.32 & 0.39 \\
        & B & 0.30 & \multicolumn{2}{c}{\em idem} & 18.84 & 18.67 & \multicolumn{2}{c}{\em idem} & 14.08 & -20.05 & 21.29 & 21.65 & 21.2 & 13.05 & -21.08 & 0.39 \\
     94 & R & 0.30 & 0.14 & 0.7 &                   21.43 & 21.33 & 3.7  & 2.5                   & 15.58 & -18.50 & 20.04 & 20.42 & 11.9 & 13.00 & -21.08 & 0.09 \\
        & B & 0.30 & \multicolumn{2}{c}{\em idem} & 23.24 & 23.08 & \multicolumn{2}{c}{\em idem} & 17.41 & -16.74 & 21.02 & 21.33 & 12.0 & 14.01 & -20.14 & 0.04 \\
    624 & R & 0.32 & 0.33 & 0.5 &                   21.10 & 20.95 & 15.0 & 3.7                   & 12.30 & -21.91 & 21.17 & 21.92 & 18.3 & 13.34 & -20.87 & 2.60 \\
        & B & 0.32 & \multicolumn{2}{c}{\em idem} & 22.94 & 22.70 & \multicolumn{2}{c}{\em idem} & 14.19 & -20.12 & 21.87 & 22.53 & 17.7 & 14.34 & -19.97 & 1.15 \\
        & I & 0.32 & \multicolumn{2}{c}{\em idem} & 20.52 & 20.41 & \multicolumn{2}{c}{\em idem} & 11.82 & -22.35 & 19.00 & 19.79 & 12.9 & 12.26 & -21.91 & 1.50 \\
   1541 & R & 0.37 & 0.16 & 0.6 &                   19.10 & 18.95 & 4.2  & 2.2                   & 13.09 & -21.49 & 20.84 & 21.03 & 19.6 & 12.73 & -21.85 & 0.72 \\
        & B & 0.37 & \multicolumn{2}{c}{\em idem} & 20.78 & 20.54 & \multicolumn{2}{c}{\em idem} & 14.77 & -19.90 & 22.26 & 22.36 & 21.5 & 13.92 & -20.75 & 0.46 \\
        & I & 0.37 & \multicolumn{2}{c}{\em idem} & 18.40 & 18.29 & \multicolumn{2}{c}{\em idem} & 12.39 & -22.15 & 19.91 & 20.14 & 18.1 & 12.02 & -22.52 & 0.71 \\
   2487 & R & 0.33 & 0.11 & 0.7 &                   19.89 & 19.40 & 6.7  & 1.7                   & 12.94 & -21.69 & 20.35 & 20.12 & 24.4 & 11.69 & -22.94 & 0.32 \\
        & B & 0.33 & \multicolumn{2}{c}{\em idem} & 21.52 & 20.73 & \multicolumn{2}{c}{\em idem} & 14.57 & -20.37 & 22.24 & 21.71 & 27.5 & 13.40 & -21.54 & 0.34 \\
   2916 & R & 0.31 & 0.12$^\dagger$\hspace{-0.14cm} & 0.4$^\dagger$\hspace{-0.135cm} & 21.09 & 20.34 & 8.8 & 2.3 & 13.40 & -21.36 & 21.57 & 20.99 & 16.1 & 13.60 & -21.16 & 1.20 \\
        & B & 0.31 & \multicolumn{2}{c}{\em idem} & 22.97 & 21.76 & \multicolumn{2}{c}{\em idem} & 15.26 & -19.97 & 22.31 & 21.27 & 14.6 & 14.68 & -20.55 & 0.59 \\
   2953 & R & 0.073& 0.20 & 0.6 & 21.34$^{\$}$\hspace{-0.145cm} & 20.22$^{\$}$\hspace{-0.145cm} & 24.1 & 3.3 & 11.38$^{\$}$\hspace{-0.145cm} & -20.64$^{\$}$\hspace{-0.145cm} & 19.92$^{\$}$\hspace{-0.145cm} & 19.25$^{\$}$\hspace{-0.145cm} & 56.1 &  9.69$^{\$}$\hspace{-0.145cm} & -22.33$^{\$}$\hspace{-0.145cm} & 0.21 \\
        & B & 0.073& \multicolumn{2}{c}{\em idem} & 23.60 & 21.78 & \multicolumn{2}{c}{\em idem} & 13.68 & -19.03 & 21.72 & 20.36 & 53.8 & 11.68 & -21.03 & 0.16 \\
        & I & 0.073& \multicolumn{2}{c}{\em idem} & 20.19 & 19.37 & \multicolumn{2}{c}{\em idem} & 10.26 & -21.45 & 18.68 & 18.32 & 54.1 &  8.58 & -23.13 & 0.21 \\
   3205 & R & 0.24 & 0.24$^\ddagger$\hspace{-0.14cm} & 0.7$^\ddagger$\hspace{-0.135cm} & 20.06 & 18.61 & 2.2  & 0.8$^\ddagger$\hspace{-0.135cm} & 16.03 & -18.85 & 19.92 & 19.59 & 14.8 & 13.08 & -21.80 & 0.07 \\
        & B & 0.24 & \multicolumn{2}{c}{\em idem} & 22.13 & 19.80 & \multicolumn{2}{c}{\em idem} & 18.12 & -17.65 & 22.19 & 20.97 & 18.2 & 14.94 & -20.83 & 0.05 \\
        & I & 0.24 & \multicolumn{2}{c}{\em idem} & 19.65 & 18.60 & \multicolumn{2}{c}{\em idem} & 15.78 & -18.71 & 18.97 & 19.04 & 15.7 & 12.04 & -22.45 & 0.03 \\
   3546 & R & 0.13 & 0.22 & 0.6 &                   17.48 & 17.29 & 2.2  & 1.0                   & 13.40 & -18.97 & 19.15 & 19.49 & 21.1 & 11.13 & -21.24 & 0.12 \\
        & B & 0.13 & \multicolumn{2}{c}{\em idem} & 19.14 & 18.84 & \multicolumn{2}{c}{\em idem} & 15.07 & -17.42 & 20.71 & 20.93 & 22.0 & 12.56 & -19.93 & 0.10 \\
   3580 & R & 0.093& 0.28 & 0.6 &                   20.22 & 20.08 & 8.0  & 1.4                   & 13.22 & -18.34 & 20.93 & 21.58 & 25.5 & 12.65 & -18.91 & 0.59 \\
        & B & 0.093& \multicolumn{2}{c}{\em idem} & 21.36 & 21.13 & \multicolumn{2}{c}{\em idem} & 14.37 & -17.27 & 22.25 & 22.81 & 26.7 & 13.87 & -17.77 & 0.63 \\
   3993 & R & 0.30 & 0.05$^\#$\hspace{-0.23cm} & 0.5$^\#$\hspace{-0.225cm} &         20.53 & 20.36 & 7.1  & 2.7                   & 13.16 & -20.97 & 22.45 & 22.37 & 18.2 & 14.07 & -20.06 & 2.32 \\
        & B & 0.30 & \multicolumn{2}{c}{\em idem} & 22.03 & 21.76 & \multicolumn{2}{c}{\em idem} & 14.69 & -19.54 & 23.42 & 23.23 & 24.0 & 14.91 & -19.32 & 1.23 \\
   4458 & R & 0.31 & 0.10 & 0.4 &                   19.21 & 19.12 & 5.8  & 2.6                   & 12.32 & -21.81 & 21.21 & 21.26 & 27.7 & 12.21 & -21.92 & 0.91 \\
        & B & 0.31 & \multicolumn{2}{c}{\em idem} & 20.72 & 20.57 & \multicolumn{2}{c}{\em idem} & 13.82 & -20.37 & 22.62 & 22.62 & 31.3 & 13.35 & -20.84 & 0.65 \\
   5253 & R & 0.10 & 0.15 & 0.4 &                   20.48 & 20.41 & 25.1 & 3.9                   & 10.28 & -21.41 & 21.16 & 21.32 & 52.2 & 10.80 & -20.89 & 1.62 \\
        & B & 0.10 & \multicolumn{2}{c}{\em idem} & 21.98$^{\$}$\hspace{-0.145cm} & 21.86$^{\$}$\hspace{-0.145cm} & \multicolumn{2}{c}{\em idem} & 11.82$^{\$}$\hspace{-0.145cm} & -19.92$^{\$}$\hspace{-0.145cm} & 21.95$^{\$}$\hspace{-0.145cm} & 22.06$^{\$}$\hspace{-0.145cm} & 57.6 & 11.43$^{\$}$\hspace{-0.145cm} & -20.31$^{\$}$\hspace{-0.145cm} & 0.70 \\
   6786 & R & 0.13 & 0.21 & 0.8$^*$\hspace{-0.145cm} & 22.43 & 22.35 & 43.5 & 5.5                & 11.18 & -20.97 & 18.27$^{@}$\hspace{-0.205cm} & 19.30$^{@}$\hspace{-0.21cm} & 12.3$^{@}$\hspace{-0.21cm} & 12.66 & -19.49 & 3.91 \\
   6787 & R & 0.092& 0.38 & 0.4$^{**}$\hspace{-0.28cm} & 18.43 & 18.37 & 8.8  & 2.3              & 11.13 & -20.31 & 19.76 & 20.49 & 36.2 & 10.70 & -20.74 & 0.68 \\
        & B & 0.092& \multicolumn{2}{c}{\em idem} & 19.89 & 19.80 & \multicolumn{2}{c}{\em idem} & 12.59 & -18.88 & 21.30 & 22.00 & 42.9 & 11.92 & -19.55 & 0.54 \\
   8699 & R & 0.18 & 0.40 & 0.6 &                   19.58 & 19.55 & 6.9  & 2.7                   & 12.77 & -20.08 & 20.65 & 22.24 & 20.8 & 13.14 & -19.71 & 1.41 \\
        & B & 0.18 & \multicolumn{2}{c}{\em idem} & 21.14 & 21.10 & \multicolumn{2}{c}{\em idem} & 14.29 & -18.58 & 21.81 & 23.39 & 21.0 & 14.13 & -18.74 & 0.86 \\
     \hline
   \multicolumn{10}{l}{$^\dagger$Galaxy with nuclear bar. Ellipticity of bulge isophotes does not converge} & 
             \multicolumn{7}{l}{$^\ddagger$Value poorly constrained due to small size of bulge. See text.} \\
   \multicolumn{10}{l}{\hspace{0.06cm} to constant value; average value is ad hoc estimate only.} & 
             \multicolumn{7}{l}{$^\$$ Values from non-photometric data.} \\
   \multicolumn{10}{l}{$^\#$Values poorly constrained due to small inclination angle of galaxy. See text.} & 
             \multicolumn{7}{l}{$^@$ Not a regular exponential disk. See text.} \\ 
   \multicolumn{10}{l}{$^*$Bulge axis ratio calculated using inclination angle of 69\deg. See text.} & 
             \multicolumn{7}{l}{} \\ 
   \multicolumn{10}{l}{$^{**}$Bulge axis ratio calculated using inclination angle of 61\deg. See text.} & 
             \multicolumn{7}{l}{} 
    \end{tabular} 
  \end{rotate}
 \end{minipage}
\end{table*}

\begin{table*}
 \begin{minipage}{15cm}
 \vspace{17cm}
 \centering
  \rotcaption{Table~\ref{table:BDresults} -- {\em continued}} \hspace{2.65cm}
  \begin{rotate}{90}
   \centering
    \hspace{0cm}%
    \begin{tabular}{rcd{1.2}@{\hspace{0.7cm}}rrrrrrrr@{\hspace{0.7cm}}rrrrr@{\hspace{0.7cm}}r}  
     \hline 
     \multicolumn{1}{c}{UGC} & band &
     \multicolumn{1}{c@{\hspace{0.7cm}}}{scale} & 
     \multicolumn{8}{c@{\hspace{0.7cm}}}{bulge parameters} &  
     \multicolumn{5}{c@{\hspace{0.7cm}}}{disk parameters} & \\

      & & & \multicolumn{1}{c}{$\epsilon_b$} &
     \multicolumn{1}{c}{$q_b$} & \multicolumn{1}{c}{$\mu_e$} &
     \multicolumn{1}{c}{$\mu_e^c$} & \multicolumn{1}{c}{$r_e$} &
     \multicolumn{1}{c}{$n$} & \multicolumn{1}{c}{$m_b$} & 
     \multicolumn{1}{c@{\hspace{0.7cm}}}{$M_b$} &
     \multicolumn{1}{c}{$\mu_0$} & \multicolumn{1}{c}{$\mu_0^c$} &
     \multicolumn{1}{c}{$h$} & \multicolumn{1}{c}{$m_d$} & 
     \multicolumn{1}{c@{\hspace{0.7cm}}}{$M_d$} &
     \multicolumn{1}{c}{B/D} \\ 

      & & \multicolumn{1}{c@{\hspace{0.7cm}}}{kpc/\arcsec} & & & 
     \multicolumn{1}{c}{$\frac{\mathrm{mag}}{\mathrm{arcsec}^2}$} & 
     \multicolumn{1}{c}{$\frac{\mathrm{mag}}{\mathrm{arcsec}^2}$} & 
     \multicolumn{1}{c}{\arcsec} & & \multicolumn{1}{c}{mag} & 
     \multicolumn{1}{c@{\hspace{0.7cm}}}{mag} &   
     \multicolumn{1}{c}{$\frac{\mathrm{mag}}{\mathrm{arcsec}^2}$} &
     \multicolumn{1}{c}{$\frac{\mathrm{mag}}{\mathrm{arcsec}^2}$} &
     \multicolumn{1}{c}{\arcsec} &  
     \multicolumn{1}{c}{mag} &
     \multicolumn{1}{c@{\hspace{0.7cm}}}{mag} & \\

     \multicolumn{1}{c}{(1)} & (2) & \multicolumn{1}{c@{\hspace{0.7cm}}}{(3)} &
     \multicolumn{1}{c}{(4)} & \multicolumn{1}{c}{(5)} &
     \multicolumn{1}{c}{(6)} & \multicolumn{1}{c}{(7)} &
     \multicolumn{1}{c}{(8)} & \multicolumn{1}{c}{(9)} &
     \multicolumn{1}{c}{(10)} & 
     \multicolumn{1}{c@{\hspace{0.7cm}}}{(11)} &
     \multicolumn{1}{c}{(12)} & \multicolumn{1}{c}{(13)} &
     \multicolumn{1}{c}{(14)} & \multicolumn{1}{c}{(15)} & 
     \multicolumn{1}{c@{\hspace{0.7cm}}}{(16)} &
     \multicolumn{1}{c}{(17)} \\    
     \hline 
  9133 & R & 0.26 & 0.32 & 0.3 &                   19.79 & 19.75 & 9.9  & 2.7                   & 12.05 & -21.66 & 20.78 & 21.27 & 34.4 & 11.69 & -22.02 & 0.72 \\
       & B & 0.26 & \multicolumn{2}{c}{\em idem} & 21.52 & 21.45 & \multicolumn{2}{c}{\em idem} & 13.78 & -19.96 & 21.93 & 22.39 & 32.4 & 12.93 & -20.81 & 0.46 \\
 11670 & R & 0.062& 0.30 & 0.6 &                   18.58 & 18.00 & 5.8  & 1.6                   & 12.20 & -18.90 & 19.09 & 19.58 & 28.5 & 10.83 & -20.27 & 0.28 \\
       & B & 0.062& \multicolumn{2}{c}{\em idem} & 20.76$^{\$}$\hspace{-0.125cm} & 19.83$^{\$}$\hspace{-0.125cm} & \multicolumn{2}{c}{\em idem} & 14.39$^{\$}$\hspace{-0.125cm} & -17.06$^{\$}$\hspace{-0.125cm} & 20.69$^{\$}$\hspace{-0.125cm} & 20.82$^{\$}$\hspace{-0.125cm} & 28.3 & 12.42$^{\$}$\hspace{-0.125cm} & -19.03$^{\$}$\hspace{-0.125cm} & 0.16 \\
       & I & 0.062& \multicolumn{2}{c}{\em idem} & 17.90 & 17.48 & \multicolumn{2}{c}{\em idem} & 11.53 & -19.41 & 18.50 & 19.15 & 30.8 & 10.05 & -20.89 & 0.26 \\
 11852 & R & 0.39 & 0.25 & 0.6 &                   20.55 & 20.36 & 5.1  & 2.2                   & 14.27 & -20.43 & 20.37 & 20.74 & 11.6 & 13.66 & -21.04 & 0.57 \\
       & B & 0.39 & \multicolumn{2}{c}{\em idem} & 22.06 & 21.76 & \multicolumn{2}{c}{\em idem} & 15.78 & -19.04 & 22.04 & 22.29 & 15.3 & 14.73 & -20.09 & 0.38 \\
       & I & 0.39 & \multicolumn{2}{c}{\em idem} & 19.98 & 19.84 & \multicolumn{2}{c}{\em idem} & 14.17 & -20.48 & 20.27 & 20.69 & 15.8 & 12.77 & -21.88 & 0.28 \\
 11914 & R & 0.072& 0.08 & 0.5 &                   20.90 & 20.66 & 26.2 & 3.1                   & 10.63 & -20.47 & 20.03 & 19.91 & 36.8 & 10.38 & -20.72 & 0.80 \\
       & B & 0.072& \multicolumn{2}{c}{\em idem} & 22.29 & 21.91 & \multicolumn{2}{c}{\em idem} & 12.06 & -19.19 & 20.74 & 20.48 & 29.9 & 11.47 & -19.78 & 0.58 \\
 12043 & R & 0.075& \rdelim\}{2}{0.5cm}[\hspace{0.1cm}No bulge component present in this galaxy] &&&&&& &        & 18.96 & 19.90 & 11.3 & 12.85 & -18.26 & --   \\
       & B & 0.075&      &     &                 &       &       &                              &       &        & 19.84 & 20.68 & 11.5 & 13.68 & -17.53 & --   \\
     \hline
   \multicolumn{10}{l}{$^\$$ Values from non-photometric data.} & 
             \multicolumn{7}{l}{} 
    \end{tabular} 
  \end{rotate}
 \hspace{11.25cm}
 \end{minipage}
\end{table*}

\clearpage

\section{Atlas of images and photometric results}
\label{app:atlas}  
On the following pages, we present the results of the photometric analysis and
bulge-disk decompositions. For all galaxies, we show on the first row three
columns with photometric data: \\   
{\bf Left hand column:} \\ 
R-band image. Grayscales are logarithmic in intensity. The cross denotes the
centre; its coordinates are given in table~\ref{table:isoresults}. A bar
indicating the physical size of the galaxy is shown in the top left of the
panel. \\  
{\bf Middle column:} \\ 
{\em Top panel:} photometric profiles in the available bands. Crosses, squares
and triangles indicate the I-, R- and B-band respectively. The errorbars on
the symbols in the top right corner show the photometric errors
$\sigma_{\mathrm {phot}}$, which are listed in
table~\ref{table:observations}. \\       
{\em Bottom panel:} colour profiles. Only points with error smaller than 0.5
mag are shown. \\    
{\bf Right hand column:} \\ 
Orientation and deviations from axisymmetry. All data in this column are
measured from the R-band images. From {\em top} to {\em bottom}: 
\begin{enumerate}
 \renewcommand{\theenumi}{--}
 \item Position angle (north through east): the data points with 
  errorbars give the results from an ellipse fit to the image with
  position angle and ellipticity as free parameters. The dashed line
  gives the adopted value which was used to derive the photometric
  profiles in the middle column. It is listed in
  table~\ref{table:isoresults}.    
 \item Ellipticity: symbols and dashed line as for the position angle
  above. 
 \item Lopsidedness: data points show the amplitude of the m=1 fourier
  term of the intensity distribution measured along the ellipses,
  relative to the average intensity at the ellipse.
 \item Phase of the m=1 fourier term: data points show the position
  angle, measured north through east, where the m=1 fourier term
  reaches maximum intensity.  
 \item Bar strength or oval distortion: same as the third panel, but
  for the m=2 fourier term.   
 \item Phase of the m=2 fourier term: same as the fifth panel, but for
  one of the maxima of the m=2 fourier term. 
\end{enumerate}
\vspace{0.25cm}
The four columns in the second row contain the following results from the
bulge-disk decompositions : \\  
{\bf First column}: \\
Model R-band image of the bulge; the grayscale used is the same as in the
original image shown in the first row. \\  
{\bf Second column}: \\
{\em Top panel:} bulge photometric profiles. Symbols show the photometric
profiles measured from the images with an initial estimate for the disk
subtracted. Only points within the bulge-disk transition radius are shown. The
solid lines show the fitted S\'ersic profiles; the fitted parameters are given
in table~\ref{table:BDresults} and were used to create the model images shown
in the first column (see section~\ref{subsec:bulges} for details). \\    
{\em Bottom panel:} bulge colour profiles. Only the colours of the fitted
profiles are shown; since we used the same effective radius and S\'ersic shape
parameter for the fits in all bands, no colour gradients are fitted, and only
a global colour is derived for the bulges. \\
{\bf Third column}: \\
Residual R-band image after subtraction of the bulge component; the
grayscale used is the same as in the original image shown in the first
row. Black marks indicate regions in the original image which were
masked in an earlier stage of the reduction process
(section~\ref{sec:obs_and_redux}). All light in this image
is assumed to originate from a flat disk component. \\
{\bf Fourth column}: \\
{\em Top panel:} disk photometric profiles. Symbols show the profiles
measured from the bulge-subtracted images. Only points outside the
bulge-disk transition radius are shown. The solid lines show the 
fitted exponential profiles; the fitted parameters are given in
table~\ref{table:BDresults} (see
section~\ref{subsec:disks} for details). \\ 
{\em Bottom panel:} disk colour profiles. Only points with error
smaller than 0.5 mag are shown. \\[0.25cm]
UGC~12043 does not have a distinctive bulge component, so no
bulge-disk decomposition was made and all light was assumed to
originate from a flattened disk (see note in section~\ref{sec:notes}).  
For this galaxy, we show on page~36 only the 3 columns with photometric
data. The fitted exponential profiles for this galaxy are overplotted with
the bold lines over the photometric profiles in the top panel of the
middle column. 
\clearpage

\begin{figure*}
    {\psfig{figure=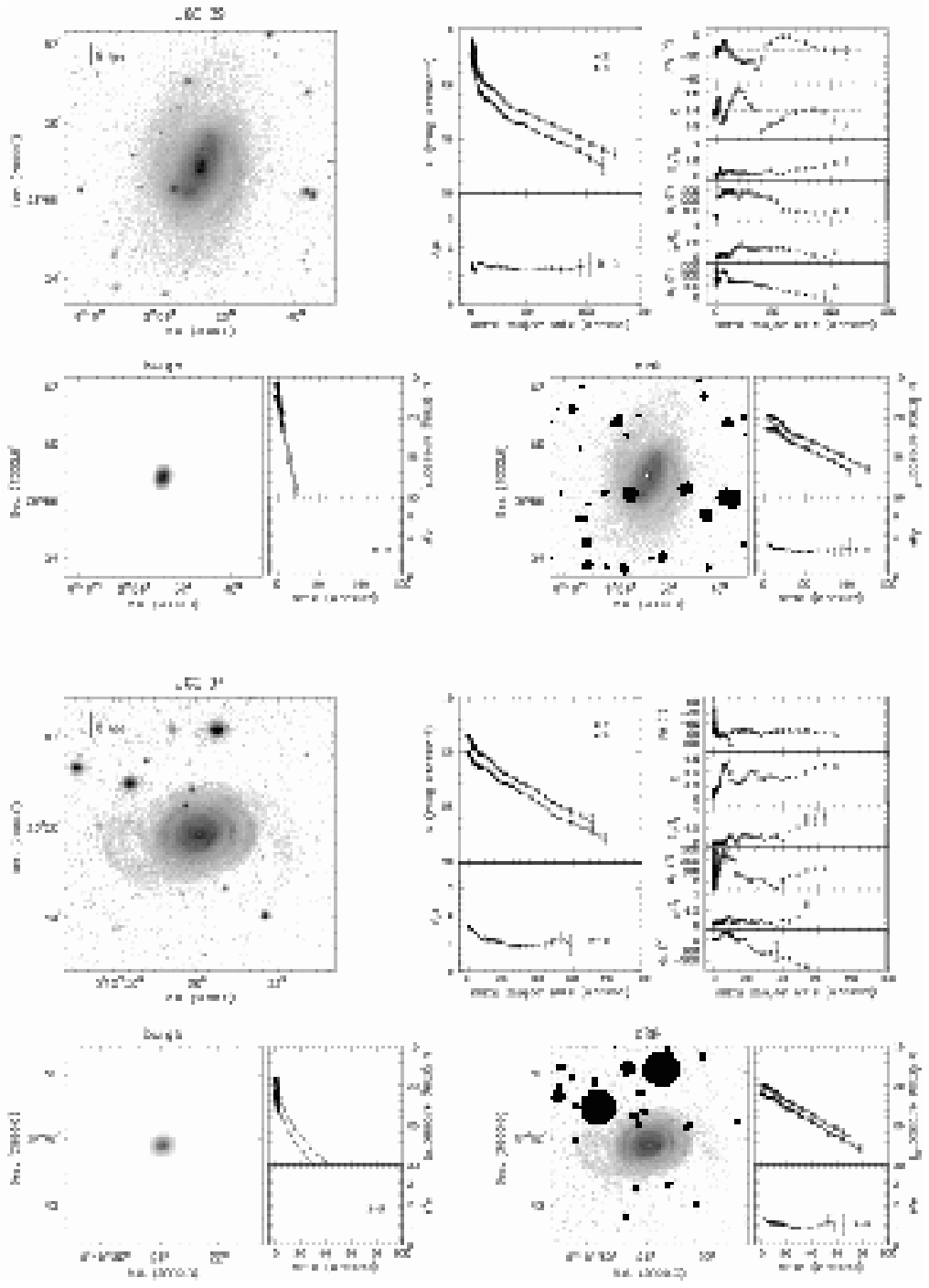,width=17.25cm}}
 \label{fig:ugc89and94}
\end{figure*}

\end{document}